\documentclass[12pt]{article}
\usepackage{amsmath}
\usepackage{graphicx,psfrag,epsf}
\usepackage{enumerate}
\usepackage{natbib}
\usepackage{textcomp}
\usepackage[hyphens]{url} 
\usepackage{hyperref}

\newcommand{\blind}{0}

\usepackage{color}
\usepackage{fancyvrb}

\DefineVerbatimEnvironment{Highlighting}{Verbatim}{commandchars=\\\{\}}
\usepackage{framed}
\definecolor{shadecolor}{RGB}{248,248,248}
\newenvironment{Shaded}{\begin{snugshade}}{\end{snugshade}}

\newcommand{\AttributeTok}[1]{\textcolor[rgb]{0.77,0.63,0.00}{#1}}

\newcommand{\CommentTok}[1]{\textcolor[rgb]{0.56,0.35,0.01}{\textit{#1}}}

\newcommand{\DocumentationTok}[1]{\textcolor[rgb]{0.56,0.35,0.01}{\textbf{\textit{#1}}}}

\newcommand{\FunctionTok}[1]{\textcolor[rgb]{0.00,0.00,0.00}{#1}}

\newcommand{\NormalTok}[1]{#1}

\newcommand{\OtherTok}[1]{\textcolor[rgb]{0.56,0.35,0.01}{#1}}

\newcommand{\StringTok}[1]{\textcolor[rgb]{0.31,0.60,0.02}{#1}}

\providecommand{\tightlist}{%
  \setlength{\itemsep}{0pt}\setlength{\parskip}{0pt}}

\usepackage{longtable,booktabs,array}
\usepackage{calc} 
\usepackage{etoolbox}
\makeatletter
\patchcmd\longtable{\par}{\if@noskipsec\mbox{}\fi\par}{}{}
\makeatother
\IfFileExists{footnotehyper.sty}{\usepackage{footnotehyper}}{\usepackage{footnote}}
\makesavenoteenv{longtable}

\usepackage[export]{adjustbox}
\usepackage{amssymb}
\usepackage{booktabs}
\usepackage{longtable}
\usepackage{array}
\usepackage{multirow}
\usepackage{wrapfig}
\usepackage{float}
\usepackage{colortbl}
\usepackage{pdflscape}
\usepackage{tabu}
\usepackage{threeparttable}
\usepackage{threeparttablex}
\usepackage[normalem]{ulem}
\usepackage{makecell}
\usepackage{xcolor}

\begin{document}

\def\spacingset#1{\renewcommand{\baselinestretch}%
{#1}\small\normalsize} \spacingset{1}


\if0\blind
{
  \title{\bf Interactive Exploration of Large Dendrograms with Prototypes}

  \author{
        Andee Kaplan \\
    Department of Statistics, Colorado State University\\
     and \\     Jacob Bien \\
    Department of Data Sciences and Operations, University of Southern California\\
      }
  \maketitle
} \fi

\if1\blind
{
  \bigskip
  \bigskip
  \bigskip
  \begin{center}
    {\LARGE\bf Interactive Exploration of Large Dendrograms with Prototypes}
  \end{center}
  \medskip
} \fi

\bigskip
\begin{abstract}
Hierarchical clustering is one of the standard methods taught for identifying and exploring the underlying structures that may be present within a data set. Students are shown examples in which the dendrogram, a visual representation of the hierarchical clustering, reveals a clear clustering structure. However, in practice, data analysts today frequently encounter data sets whose large scale undermines the usefulness of the dendrogram as a visualization tool. Densely packed branches obscure structure, and overlapping labels are impossible to read. In this paper we present a new workflow for performing hierarchical clustering via the R package called \texttt{protoshiny} that aims to restore hierarchical clustering to its former role of being an effective and versatile visualization tool. Our proposal leverages interactivity combined with the ability to label internal nodes in a dendrogram with a representative data point (called a \emph{prototype}). After presenting the workflow, we provide three case studies to demonstrate its utility.
\end{abstract}

\noindent%
\vfill

\newpage
\spacingset{1.45} 

\nocite{devtools}

\hypertarget{introduction}{%
\section{Introduction}\label{introduction}}

Clustering is one of the principal tools used by data analysts for uncovering the structure present within a data set. Hierarchical clustering is particularly popular since it can reveal multiple scales of groupings at once without forcing the data analyst to commit to a certain number of clusters. Hierarchical clustering has been used successfully in a wide range of application domains, from biology \citep{ao2005clustag, sorlie2003repeated} to social sciences \citep{kigerl2020behind, saint2003convergence} to document recovery \citep{zhao2005hierarchical, cutting2017scatter} and beyond \citep{studdert1974balance}.

\begin{figure}[ht!]
\includegraphics[width=0.5\linewidth]{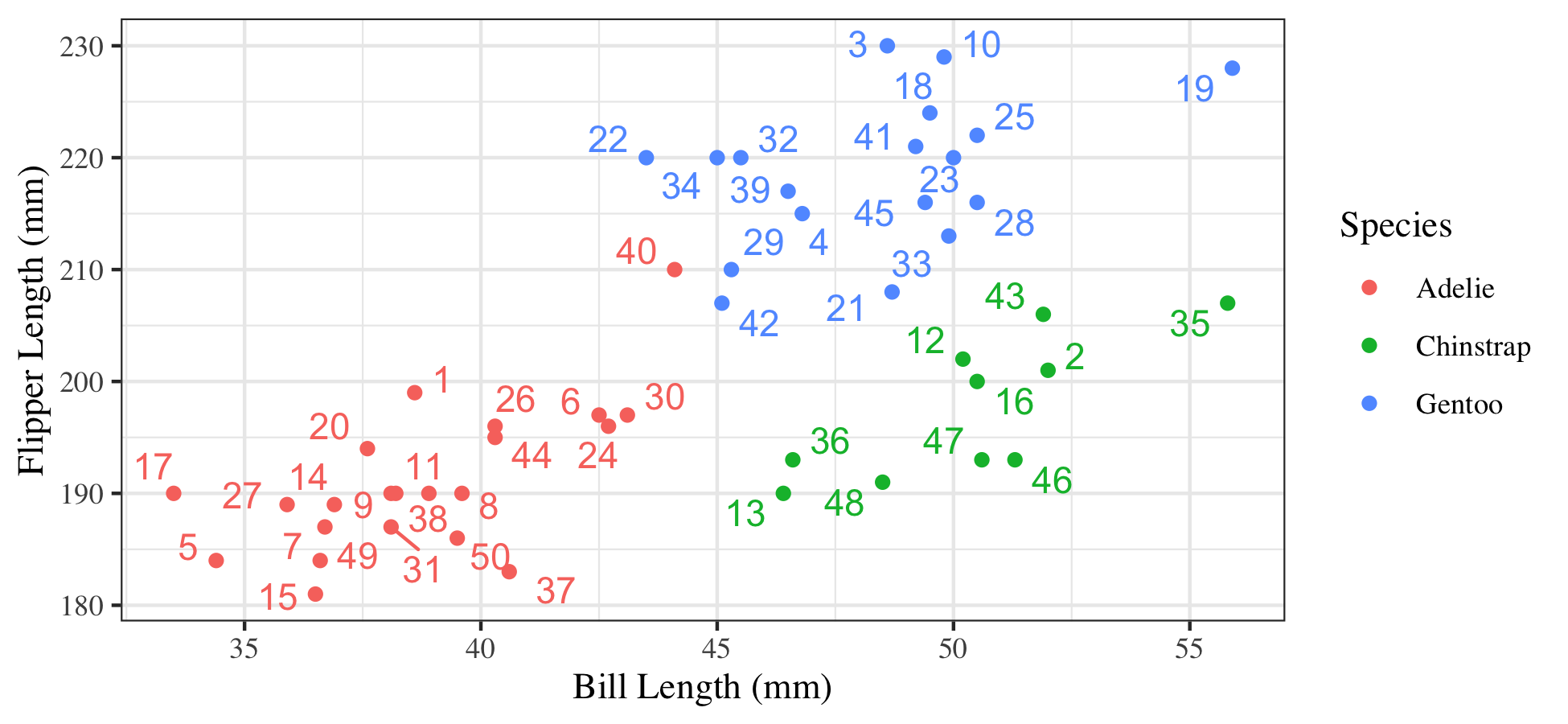} \includegraphics[width=0.5\linewidth]{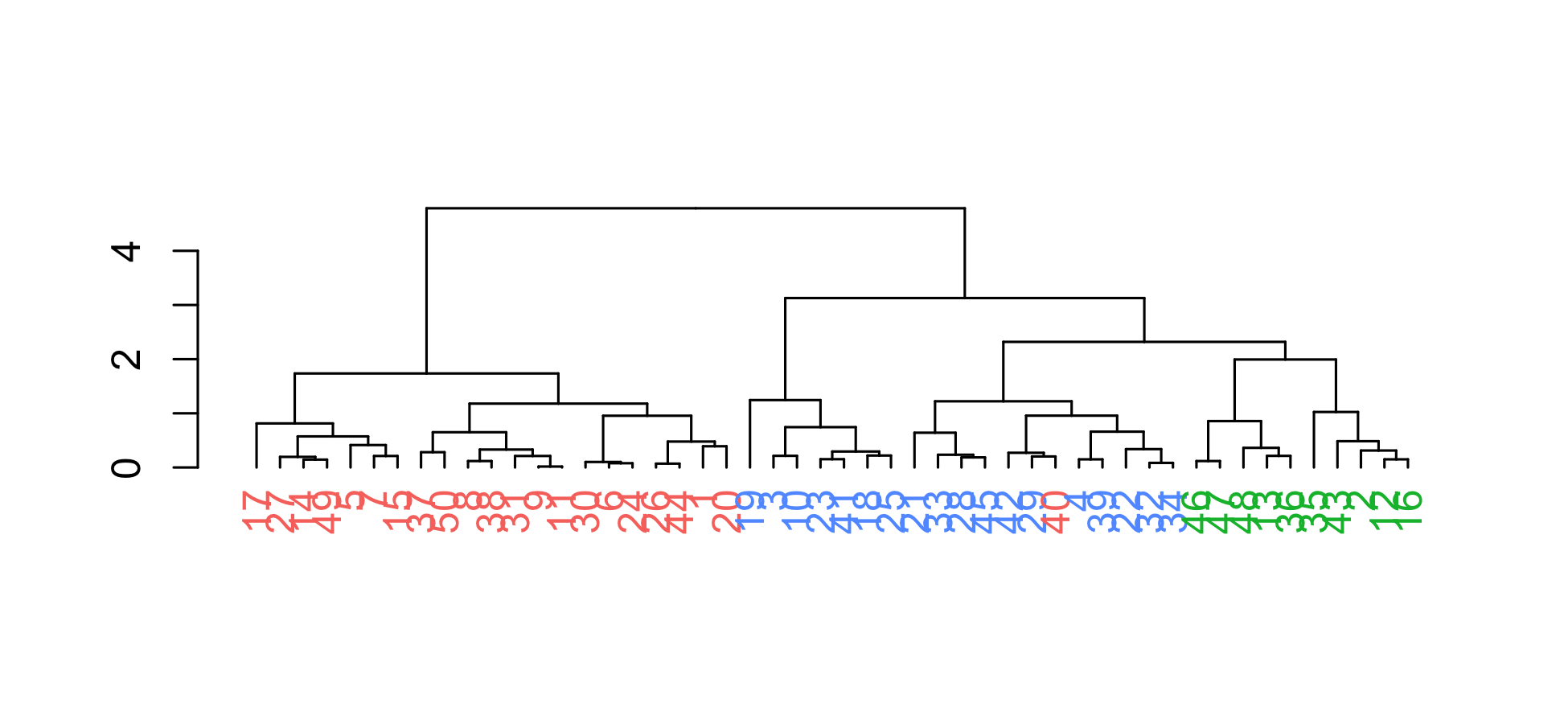} \caption{(Left) Fifty randomly selected observations of penguins' flipper and bill lengths colored by species. There appear to be three clusters that roughly correspond to the species. (Right) A dendrogram of that same data reveals the clustering structure.}\label{fig:small-dendrogram}
\end{figure}

The hierarchical clustering of a data set is represented by a dendrogram, which displays the original observations as leaves of a tree, with interior nodes of the tree corresponding to successive ``mergings'' of these observations into ever larger clusters. For example, the dendrogram in Figure \ref{fig:small-dendrogram} shows a sample of 50 observations of penguin measurements \citep{palmerpenguins}. According to the scatterplot showing bill size and flipper size, there appear to be three primary clusters that roughly correspond to the species of penguin. This is supported by the dendrogram presented.

\begin{figure}[ht!]

{\centering \includegraphics[width=5.5in]{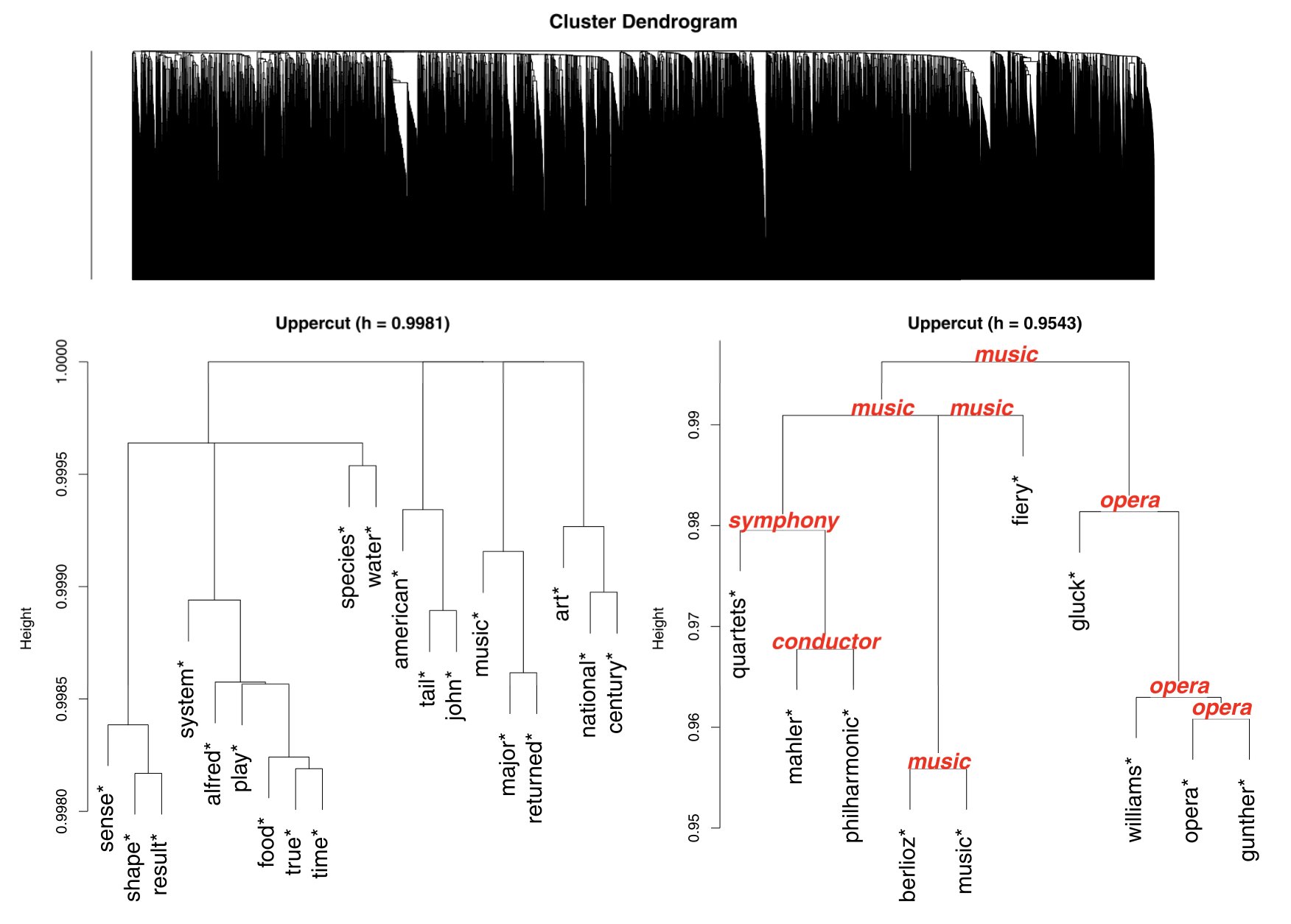} 

}

\caption{Reproduction of Figure 6 from \cite{bien2011hierarchical} showing the process of using static images to ``drill down" into a dendrogram via the use of prototypes.}\label{fig:fig-6-repro}
\end{figure}

Such is the promise of hierarchical clustering as presented in most statistics classes. Yet, despite the appeal of hierarchical clustering in such examples, its use in real applications can be hampered by practical challenges. First, its usefulness as a visualization tool is severely degraded by increasing data set sizes.
The top panel of Figure \ref{fig:fig-6-repro} shows a dendrogram for a hierarchical clustering of about \(15,000\) of the most common words from Grolier's Encyclopedia \citep{groliers}. In this dendrogram, the branches are more tightly packed, rendering the leaf labels useless due to overlap. This is a known challenge in the visualization literature called overplotting \citep{swayne1998xgobi}, where often the number of elements to be plotted exceeds the number of pixels available to create plots. A number of solutions have been proposed to address this limitation in different plot types, including the introduction of transparency \citep{cottam2013overplotting}, binning, stacking \citep{dang2010stacking}, and interactivity \citep{swayne1998xgobi}. Second, one must avoid uncritically accepting the structure revealed by a hierarchical clustering since it has been suggested that when no true clustering structure is present in a high-dimensional data set, the dendrogram can still misleadingly indicate structure that is a reflection of the the clustering method rather than the data set \citep{thrun2021distance}. This underscores the importance of being able to inspect dendrograms to understand the reasonableness of the findings based on domain expertise. Both of these practical challenges---the problem of overplotting and the need to carefully probe the recovered structure---can be alleviated by the approach described in this work, which adds interactivity into dendrograms.

While still a developing field, interactive statistical graphics has been a topic of interest since at least the late 1960s \citep{asavideolibrary, friedman2002john} and has seen emerging popularity and success in advancing exploratory data analysis \citep[e.g.,][]{tukey1977exploratory, unwin1996interactive, theus2008interactive, young2011visual, su2017glimma}. Recent development of JavaScript frameworks has made it much easier for statisticians to incorporate interactivity into statistical graphics, specifically for the web browser \citep[e.g.,][]{bostock2011d3, plotly, ggvis, animint, satyanarayan2016vega}. While much of this work has focused on the general interactivity tasks of linking plots, brushing, labeling, and scaling \citep{Swayne1999}, other work has attempted to solve more specific problems through the use of interactive statistical web graphics \citep[e.g.,][among others]{sievert2014ldavis, kaplan2016putting, kaplan2017interactive}. In this paper we present an example of the latter goal---solving the specific problem of exploring a large dendrogram through the use of interactive statistical web graphics.

The use of interactivity to explore dendrograms has been seen in a limited number of previous works. \citet{1016905} provide a desktop application for exploring dendrograms of gene expression data that allows for interaction with clusters, but does not allow one to explore portions of the dendrogram in isolation, which is necessary to visualize and understand very large dendrograms. \citet{sieger2017interactive} is a more recent example of an interactive dendrogram available in R that employs the canvas infrastructure to provide interactivity to the user with features and limitations are very similar to \citet{1016905}.
Conversely, \citet{collapsibleTree} provides an interactive tree diagram in R that allows for isolating pieces of the tree, but it is not specialized to hierarchical clustering dendrograms and requires the user to manually reformat the clustering object as a tree object. Additionally, this tool does not display the standard dendrogram feature of height, which indicates how far apart two clusters are when they are merged.

Our approach to this problem is built on work by \citet{bien2011hierarchical}, who proposed adding the labels of prototypes (i.e., cluster representatives) to the interior nodes of dendrograms and demonstrated through a series of static images how one could in principle use these prototypes to explore a hierarchical clustering in a top-down manner in a process they called ``drilling down.'' As a demonstration, Figure 6 from their paper (reproduced here as Figure \ref{fig:fig-6-repro}) shows several of these static images. The upper panel of the figure shows the full tree from the hierarchical clustering of about \(15,000\) of the most common words from Grolier's Encyclopedia \citep{groliers}. It is clear that overplotting obfuscates whatever structure might be present in this data. The bottom two portions of the figure show how one can use prototypes to alleviate this problem. On the lower left we see the ``upper cut'' view of the dendrogram, which is what one gets by only showing the nodes that are above a certain cut height and then replacing any branch that has been cut by a label of the prototype for that branch. Note that without having prototypes assigned to each interior node, there would not be a natural way of removing branches like this. We can ``drill down'' the tree by examining any of these branches. For example, one of the branches is labeled by the prototype ``music,'' and on the lower right we see the upper cut of the branch labeled by the ``music'' label. In this branch, ``conductor'' is a prototype for two branches that are prototyped by ``mahler'' and ``philharmonic.'' When this branch is merged with the branch prototyped by ``quartets,'' the new prototype becomes ``symphony.'' Working one's way down a dendrogram is referred to as ``drilling down.'' Despite this demonstration and discussion in \citet{bien2011hierarchical}, their \texttt{protoclust} package \citep{protoclust} does not provide the ability to create such images; furthermore, producing a series of static images does not lend itself well to data exploration.

The goal of this present work is to render hierarchical clustering useful again for visualizing and exploring data sets at scales of interest by introducing interaction with the dendrogram into a clustering workflow. Additionally, we provide a tool, which we call \texttt{protoshiny}, that enables this workflow by leveraging three basic ideas beyond the standard hierarchical clustering dendrogram:

\begin{enumerate}
\def\labelenumi{\arabic{enumi})}
\tightlist
\item
  Use cluster prototypes to summarize branches of a dendrogram.
\item
  Make dendrograms interactive. Rather than attempting to show an entire dendrogram, allow the data analyst to navigate it interactively through subtrees that can be expanded or contracted.
\item
  Enable the data analyst to quickly find clusters of interest via search functionality.
\end{enumerate}

The \texttt{protoshiny} R package is a tool for facilitating interactive dendrograms that enable fast finding of interesting clusters with large data sets (2-3) through the use of prototypes (1). While minimax linkage is the most direct way to produce hierarchical clustering with prototypes, \texttt{protoshiny} is designed to more generally accommodate any linkage so long as the user also specifies a choice of prototypes.

In Section \ref{protoshiny} we describe \texttt{protoshiny}, an R \citep{rstats} package and interactive dendrogram application. We begin with providing background on hierarchical clustering with prototypes and discuss the particular interactive elements incorporated into \texttt{protoshiny} and comparing the features of \texttt{protoshiny} to three other methods for visualizing a dendrogram. Section \ref{case-studies} presents three case studies of using \texttt{protoshiny} to explore large dendrograms with applications to movie clustering, flow cytometry, and studying patterns of COVID-19 spread across the US. Each case study emphasizes a different strength of \texttt{protoshiny}. We finish with a discussion of the current limitations of the tool, as well as potential directions for expansion in Section \ref{discussion}.

\hypertarget{protoshiny}{%
\section{Prototypes and Working with Interactivity}\label{protoshiny}}

\hypertarget{hierarchical-clustering-with-prototypes}{%
\subsection{Hierarchical Clustering with Prototypes}\label{hierarchical-clustering-with-prototypes}}

Agglomerative hierarchical clustering requires the specification of what is known as a linkage, which describes how one measures the dissimilarity between pairs of clusters. For example, suppose \(G,H\subseteq \{\boldsymbol x_1,\ldots, \boldsymbol x_n\}\) are two disjoint sets of observations and \(d\) is a measure of dissimilarity between individual observations. One of the most common linkages is complete linkage, which measures the separation between clusters \(G\) and \(H\) based on the farthest between-cluster pair of observations:
\[
d_{\text{complete}}(G,H)=\max_{\boldsymbol x\in G,\boldsymbol x'\in H}d(\boldsymbol x,\boldsymbol x').
\]
Minimax linkage \citep{ao2005clustag, bien2011hierarchical} is a newer linkage that measures cluster separation based on how well the pair of clusters can be summarized by a single observation from one of the two clusters. The key distinguishing property of minimax linkage is that it provides a natural definition of ``prototype'' for each cluster produced in the hierarchical clustering. A prototype is a representative element of the cluster that is chosen from one of the original observations. Having the prototype be one of the original observations is important for interpretability. For example, in Figure \ref{fig:fig-6-repro}, the average of a collection of vectors representing a word is far less useful than a single well-chosen word. In non-hierarchical clustering settings, the k-medoids method is used for this same reason (see, e.g., \citet{hastie2009elements}).

\begin{figure}[ht!]

{\centering \includegraphics[width=0.7\linewidth]{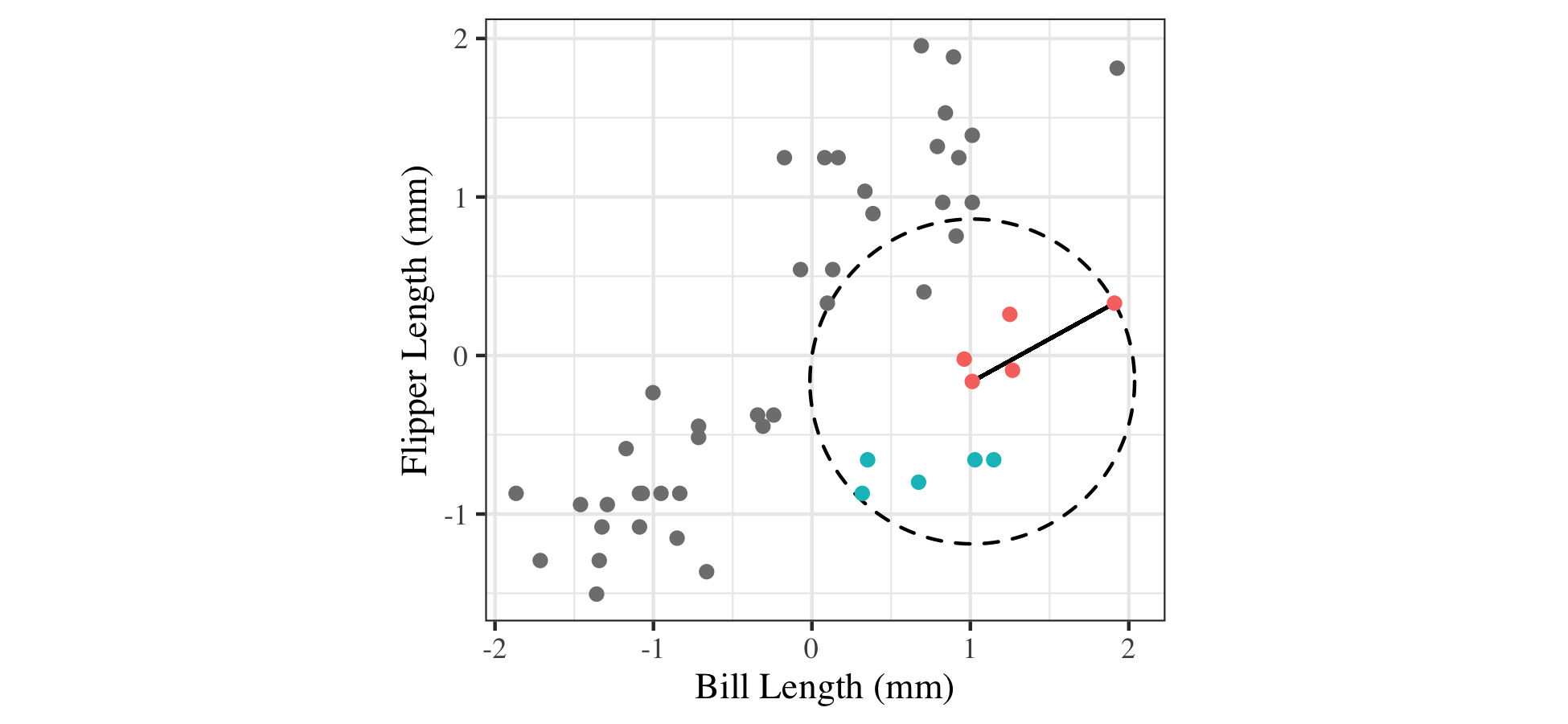} 

}

\caption{Demonstration of minimax linkage on (centered and scaled) Palmer penguins data. The solid black line represents the distance between the red and green clusters. The dotted circle is of radius $r(G \cup H)$, where $G$ and $H$ denote the two clusters, and is centered at the prototype for the cluster formed by merging $G$ and $H$.}\label{fig:minimax-demo}
\end{figure}

To describe minimax linkage, one starts by defining the dissimilarity between an observation and a set:
\[
d_{\max}(\boldsymbol x,C)=\max_{\boldsymbol x'\in C}d(\boldsymbol x,\boldsymbol x').
\]
The \emph{minimax radius} of a set \(C\) is then defined as the size of the smallest enclosing ``ball'' of \(C\) that is centered at an element of \(C\),
\[
r(C)=\min_{\boldsymbol x\in C}d_{\max}(\boldsymbol x,C).
\]
The center of this ball,
\[
p(C)=\arg\min_{\boldsymbol x\in C}d_{\max}(\boldsymbol x,C),
\]
is defined as the \emph{minimax prototype} for the set \(C\). Because the minimum is taken specifically over \(x\in C\), by definition \(p(C)\) will always be one of the elements of \(C\). If \(r(C)\le h\), then all points in \(C\) are within a dissimilarity of \(h\) of the prototype \(p(C)\). The minimax linkage between \(G\) and \(H\) is then defined as
\[
d_{\text{minimax}}(G,H)=r(G\cup H),
\]
and if clusters \(G\) and \(H\) are merged together, the newly formed cluster \(G\cup H\) has prototype \(p(G\cup H)\). A demonstration of the minimax linkage as used to merge two clusters in the (centered and scaled) Palmer penguins data is given in Figure \ref{fig:minimax-demo}.

\citet{bien2011hierarchical} showed that minimax linkage has a number of desirable properties. For example, suppose one ``cuts'' a minimax linkage dendrogram at height \(h\) to produce a set of clusters. In such a case, we are guaranteed that every point in the data set is within a dissimilarity of \(h\) of its prototype. They also discussed its efficient implementation and described how ``\,`prototype-enhanced' dendrograms'' provide a convenient way of ``drilling down'' a dendrogram (as we described in the discussion of Figure \ref{fig:fig-6-repro}). The key idea is that minimax linkage provides every interior node of a dendrogram with an associated prototype that can be used as a label for summarizing the branch of observations beneath it. This allows one to prune the dendrogram, replacing certain branches of the dendrogram by their prototypes. While \citet{bien2011hierarchical} demonstrated how one might explore a dendrogram in this fashion, a tool for facilitating such a process was not developed.

\hypertarget{interactivity}{%
\subsection{Data Exploration with Interactivity}\label{interactivity}}

We incorporate three tools of interactivity that allow the data analyst to take full advantage of having a prototype-labeled dendrogram in their analysis of hierarchically clustered data:

\begin{enumerate}
\def\labelenumi{\arabic{enumi})}
\tightlist
\item
  expansion/contraction (drill down),
\item
  zooming and panning, and
\item
  search functionality.
\end{enumerate}

Expansion and contraction of nodes in the dendrogram is key for carrying out the drilling down process described in the discussion of Figure \ref{fig:fig-6-repro}. The data analyst can choose which parts of the tree to see in detail and which to hide from view. A potential workflow is as follows. One starts with an upper cut view (analogous to the lower left panel of Figure \ref{fig:fig-6-repro}) and uses the visible prototype labels as a high-level summary that suggests where to further explore. The zooming and panning of the dendrogram can be used to focus attention on a particular portion of the dendrogram that may have become crowded due to a large number of expanded elements. In a second potential workflow, the search functionality allows a data analyst to find the first (i.e., highest) instance of a label in the dendrogram. This is useful for quickly locating a cluster with a label that is of interest \emph{a priori} to the data analyst.

\hypertarget{additional-details-and-usage}{%
\subsection{Additional Details and Usage}\label{additional-details-and-usage}}

The interactive browser-based application \texttt{protoshiny} is built using the \texttt{Shiny} framework \citep{shiny} and the JavaScript library \texttt{D3} \citep{bostock2011d3}. It is an open-source application and available at \url{https://github.com/andeek/protoshiny}. In order to use \texttt{protoshiny}, a user can install the R package and launch the application with the following commands:

\begin{Shaded}
\begin{Highlighting}[]
\DocumentationTok{\#\# install package}
\FunctionTok{install.packages}\NormalTok{(}\StringTok{"protoshiny"}\NormalTok{)}

\DocumentationTok{\#\# launch application}
\FunctionTok{library}\NormalTok{(protoshiny)}
\FunctionTok{visualize\_hc}\NormalTok{()}
\end{Highlighting}
\end{Shaded}

The application is launched in a web browser window, and users can interact with \texttt{protoshiny} by either uploading their own \texttt{protoclust} object (the result of running hierarchical clustering with minimax linkage) or using one of the default test data sets that are pre-loaded. The \texttt{protoshiny} package contains a convenience function (\texttt{as.protoclust}) for converting general clustering objects to \texttt{protoclust} objects with the addition of a user-specified vector of prototypes.

\texttt{protoshiny} is most useful in situations where the labels of clustered objects have interpretable meaning. Here, ``label'' can be either some text or a thumbnail image. Table \ref{tab:compare} provides a comparison of functionality between \texttt{protoshiny} and three other options -- two interactive and one static. The inclusion of interactivity, like collapsible nodes, zoom and pan, and search functionality, in combination with labelled branches contribute to the utility of \texttt{protoshiny} when compared to other options. Furthermore, the web-based framework allows for a hosted version to remove any need for a user to install software. \citet{collapsibleTree} also provide a web-based interactive tree, however this package does not display the tree structure as a dendrogram, thus losing the visual representation of similarity between elements through height of branches. \citet{sieger2017interactive} provides improvements over a static dendrogram in situations where a heatmap is useful whereas \texttt{protoshiny} is most useful when observation labels are informative (e.g., clustering words, movies, images, counties). In some cases, a dissimilarity is all that can be computed (without observations being points in a space) and in such a situation a heatmap is not even available.

\begin{table}

\caption{\label{tab:unnamed-chunk-2}\label{tab:compare} Comparison of functionality between \texttt{protoshiny} and three other options -- two interactive and one static.}
\centering
\begin{tabular}[t]{l>{\centering\arraybackslash}p{1in}>{\centering\arraybackslash}p{1in}>{\centering\arraybackslash}p{1in}>{\centering\arraybackslash}p{1.2in}}
\toprule
Functionality & $\texttt{protoshiny}$ & Static Dendrogram & $\texttt{idendr0}$ (Sieger et al. 2017) & $\texttt{collapsibleTree}$ (Khan 2018)\\
\midrule
\cellcolor{gray!6}{Interactivity} & \cellcolor{gray!6}{$\checkmark$} & \cellcolor{gray!6}{} & \cellcolor{gray!6}{$\checkmark$} & \cellcolor{gray!6}{$\checkmark$}\\
Zoom and Pan & $\checkmark$ &  & $\checkmark$ & $\checkmark$\\
\cellcolor{gray!6}{Tree as Dendrogram} & \cellcolor{gray!6}{$\checkmark$} & \cellcolor{gray!6}{$\checkmark$} & \cellcolor{gray!6}{$\checkmark$} & \cellcolor{gray!6}{}\\
Cluster Export & $\checkmark$ &  & $\checkmark$ & \\
\cellcolor{gray!6}{Large Data} & \cellcolor{gray!6}{$\checkmark$} & \cellcolor{gray!6}{} & \cellcolor{gray!6}{$\checkmark$} & \cellcolor{gray!6}{}\\
\addlinespace
Web-based & $\checkmark$ &  &  & $\checkmark$\\
\cellcolor{gray!6}{Labelled Branches} & \cellcolor{gray!6}{$\checkmark$} & \cellcolor{gray!6}{} & \cellcolor{gray!6}{} & \cellcolor{gray!6}{$\checkmark$}\\
Collapsible Nodes & $\checkmark$ &  &  & $\checkmark$\\
\cellcolor{gray!6}{Thumbnail Images} & \cellcolor{gray!6}{$\checkmark$} & \cellcolor{gray!6}{} & \cellcolor{gray!6}{} & \cellcolor{gray!6}{}\\
Search & $\checkmark$ &  &  & \\
\addlinespace
\cellcolor{gray!6}{Heatmap} & \cellcolor{gray!6}{} & \cellcolor{gray!6}{} & \cellcolor{gray!6}{$\checkmark$} & \cellcolor{gray!6}{}\\
Linked Brushing &  &  & $\checkmark$ & \\
\bottomrule
\end{tabular}
\end{table}

\hypertarget{case-studies}{%
\section{Case Studies}\label{case-studies}}

We now demonstrate the workflow through three case studies. These examples highlight various features of the \texttt{protoshiny} tool and convey how they can lead to greater insight into a data set.

\hypertarget{movies}{%
\subsection{Movies}\label{movies}}

In this section, we explore a hierarchical clustering of 13,816 movies. We use the \emph{MovieLens 25M Data set} \citep{harper2015movielens},
which is based on users' ratings and taggings of movies. \citet{vig2012tag} show how movies can be embedded in a vector space in which each dimension gives the relevance score of this movie to a particular tag. In the data we use, there are 1,128 such tags. Each movie is represented by a 1,128-dimensional vector of relevance scores. For example, the five tags with the highest relevance scores for the 1993 movie \emph{Groundhog Day} are \texttt{time\ loop}, \texttt{comedy}, \texttt{original}, \texttt{imdb\ top\ 250}, and \texttt{small\ town}.

Given this embedding, we perform hierarchical clustering with minimax linkage and dissimilarities given by one minus the correlation between the movies' relevance score vectors. An example of creating and saving a \texttt{protoclust} object for this clustering is provided below. In the following code snippet, \texttt{D} represents a matrix of the dissimilarities between movie vector embeddings. After the cluster object is created, the \texttt{protoshiny} application can be launched to visually explore the results.

\begin{Shaded}
\begin{Highlighting}[]
\FunctionTok{library}\NormalTok{(protoclust) }\CommentTok{\# clustering with minimax linkage}

\DocumentationTok{\#\# perform clustering on a distance matrix D}
\NormalTok{hc }\OtherTok{\textless{}{-}} \FunctionTok{protoclust}\NormalTok{(D)}

\DocumentationTok{\#\# save object in known location}
\FunctionTok{save}\NormalTok{(hc, }\AttributeTok{file =} \StringTok{"directory/hc.Rdata"}\NormalTok{)}
\end{Highlighting}
\end{Shaded}

\begin{figure}[ht!]
\includegraphics[width=6.5in]{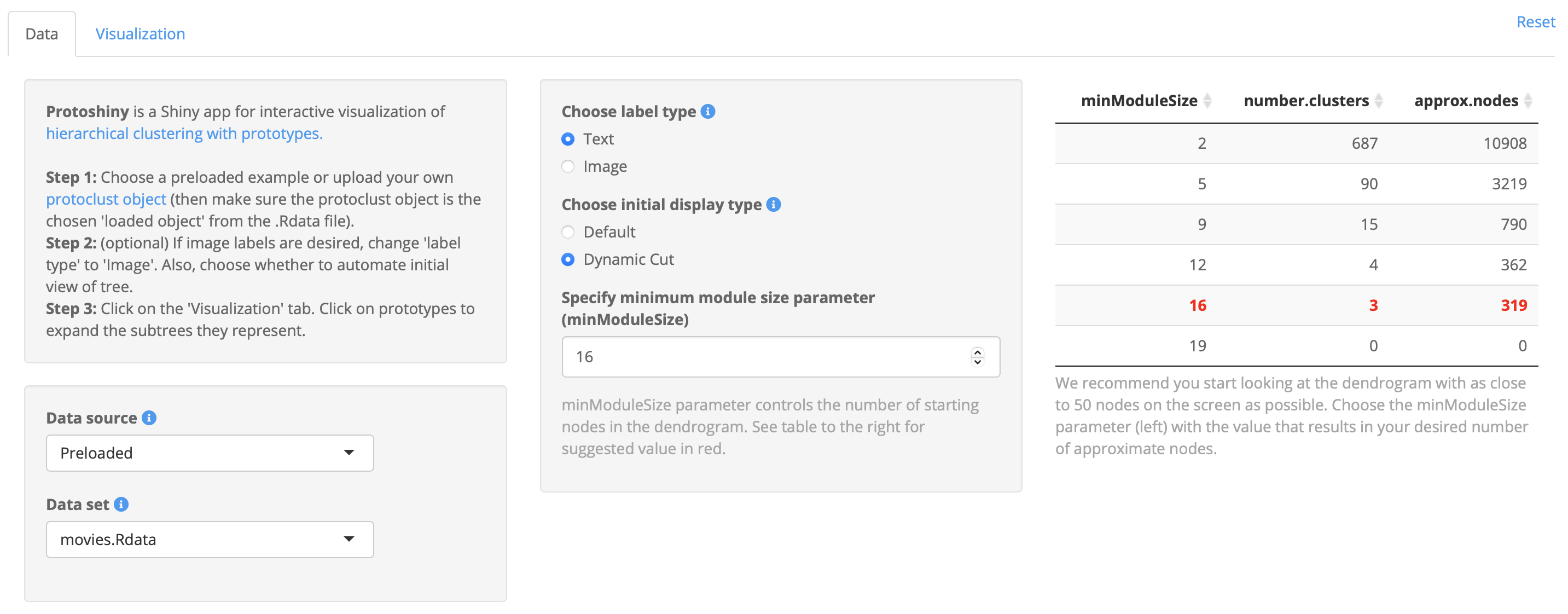} \caption{A screenshot of the initialization screen in \texttt{protoshiny}. A user has a choice of labels used in the dendrogram (text or image) as well as an initial state of the dendrogram (top 10 nodes or dynamic cut).}\label{fig:initial}
\end{figure}

Once the data is loaded into the application via the interface, the user can choose for prototype labels to be either text (in which case labels come from the row names of \texttt{D}) or thumbnail images. A screen capture of the option specification tab of the application is shown in Figure \ref{fig:initial}. For the movie example, we will select text labels. There are also two choices for the initial state of the dendrogram. The default is to show the top 10 nodes. A second option, called ``Dynamic Cut'' is also included. In traditional usage, hierarchical clustering yields a choice of \(n-1\) different clusterings, one clustering for each step of the algorithm. Each interior node of a dendrogram represents the merging of a pair of clusters, and the height of that node has meaning----the height of the interior node corresponding to the merging of clusters \(G\) and \(H\) is given by \(d(G,H)\), where \(d\) is the particular linkage being used. Traditionally, to get a particular clustering, one ``cuts'' the dendrogram at a chosen height \(h\), returning the clustering given by the \(k\) branches resulting from the cut.

\citet{langfelder2007defining} introduce an alternative to the fixed height cutting method for dendrograms that is intended to have improved performance on complex dendrograms, called ``Dynamic Cut.'' The proposed dynamic method is an adaptive approach that starts with a fixed height cut and then iteratively splits and combines clusters until the number of clusters becomes stable. The joining heights of the initial clusters are used to detect sub-cluster structure via a run-based calibration procedure. If sub-cluster structure is detected, this cluster is split. Clusters are merged when their membership becomes too small. The idea being that dynamic tree cutting may produce a more suitable starting view of the dendrogram in an automated fashion. We have incorporated the dynamic tree cutting methods in \texttt{protoshiny} by using the R package \texttt{dynamicTreeCut} \citep{dynamicTreeCut}. In \texttt{protoshiny}, the user has the ability to specify the minimum size of the final clusters resulting from \texttt{dynamicTreeCut}, which directly affects the number of nodes seen in the initial view of the dendrogram. For visualization, we recommend choosing a minimum size parameter that results in approximately 50 nodes to be displayed, however this choice can be manually adjusted by the user.

Once all options have been set, the data analyst can view and interact with the dendrogram by clicking the ``Visualization'' tab in the application. At this point an initial dendrogram will be displayed
and the data analyst can interact with the dendrogram in the three ways described in Section \ref{interactivity}: by clicking on nodes to expand/contract them; by zooming and panning to particular portions of the dendrogram using the scroll or click and drag functionality of the analyst's mouse; and lastly, by use of the search bar to reveal the first instance of a particular prototype label. Prototype labels will be shown on each branch of the dendrogram only when they differ from the parent branch prototype.

\begin{figure}[ht!]
\includegraphics[width=.5\textwidth,fbox]{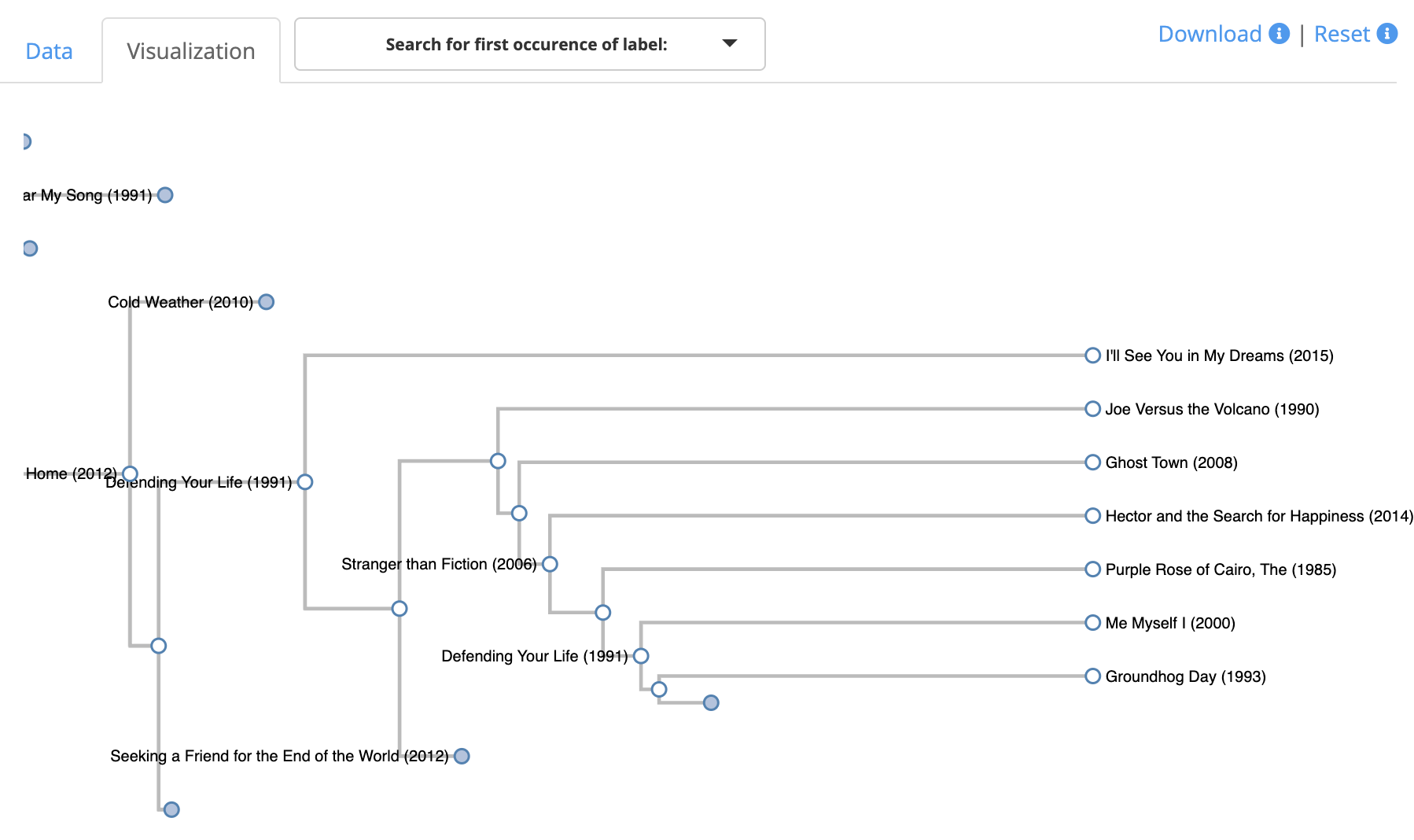} \includegraphics[width=.5\textwidth,fbox]{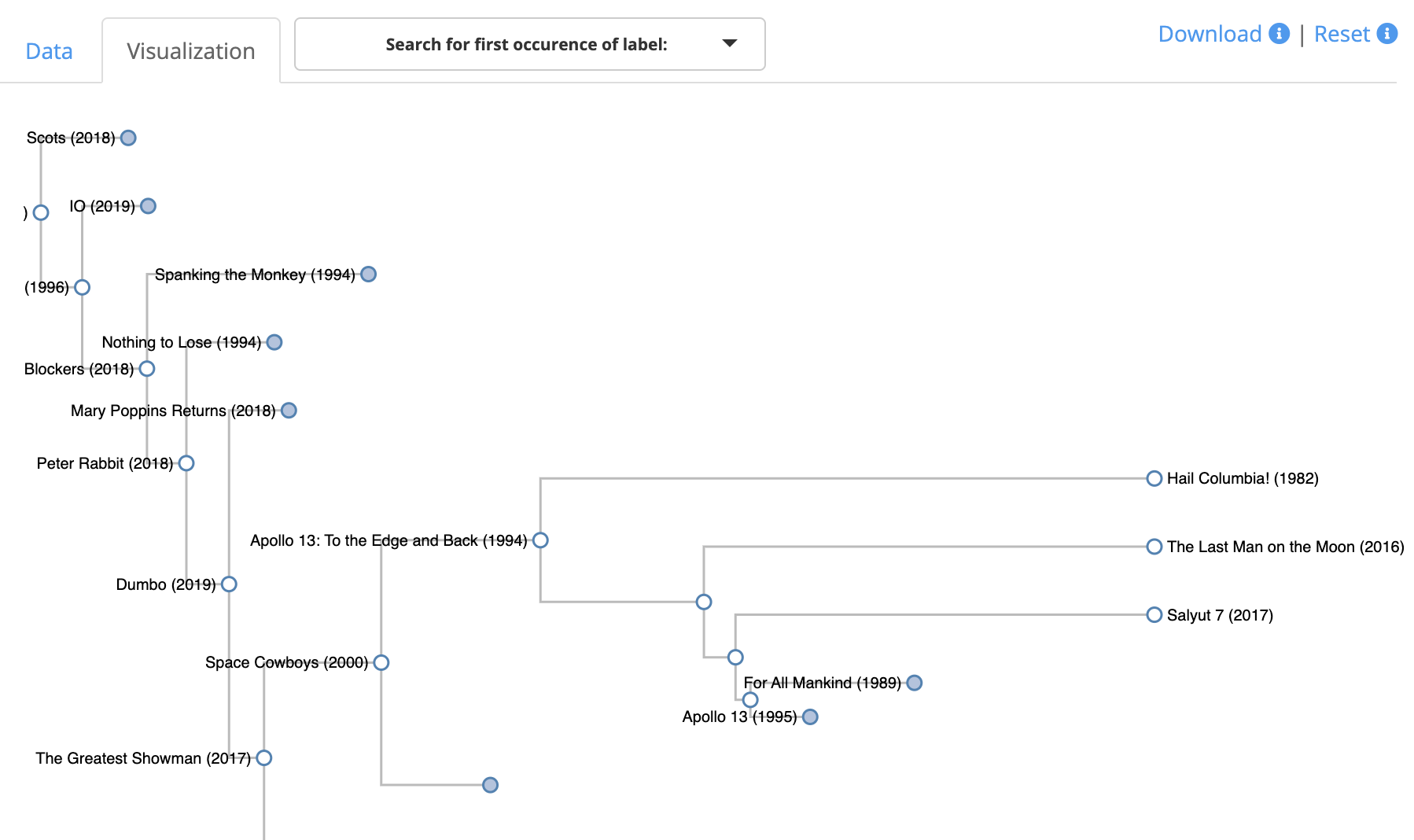} \caption{(Left) A screenshot of the movies dendrogram after performing a search for the movie {\em Groundhog Day}. (Right) A screenshot of the movies dendrogram after performing a search for the movie {\em Apollo 13}}\label{fig:groundhogday}
\end{figure}

The left side of Figure \ref{fig:groundhogday} shows the result of using the search feature to find \emph{Groundhog Day} in the dendrogram. We see that its lowest prototypes in the tree are \emph{Defending Your Life} and \emph{Stranger than Fiction}. These movies are natural choices as prototypes for \emph{Groundhog Day} since they all are a combination of comedy, drama, and fantasy. The search feature finds the highest occurrence in the dendrogram of a movie. In the case of \emph{Groundhog Day}, it is not a prototype for any movie and therefore the search revealed the movie as a leaf in the dendrogram, which is the highest occurrence of this movie in the dendrogram. However, the right side of Figure \ref{fig:groundhogday} shows that when we search for the 1995 movie \emph{Apollo 13}, this movie is a prototype for a branch of the tree, hence it shows up higher in the dendrogram. Expanding this branch of the tree reveals that it is a prototype for four historical space-related movies. The search feature returns only the highest occurrence of a label in the dendrogram. For example, if the branch shown for \emph{Apollo 13} contained a child branch for which \emph{Apollo 13} was also a prototype, it would not be displayed initially.

We provide a comparison of dendrograms resulting from a static plot, \texttt{idendr0} \citep{sieger2017interactive}, and \texttt{collapsibleTree} \citep{collapsibleTree} in the supplementary material for this case study.

\hypertarget{flow-cytometry-in-the-ocean}{%
\subsection{Flow Cytometry in the Ocean}\label{flow-cytometry-in-the-ocean}}

To study the time-changing biogeography of phytoplankton, oceanographers have developed the ability to perform continuous-time flow cytometry measurements on a ship as it travels through the ocean \citep{swalwell2011seaflow}. The output is a sequence of three-dimensional scatterplots, in which points represent individual cells and a point's location in the scatterplot describes three optical properties of that cell. In this section, we use \texttt{protoshiny} to explore the data collected from a cruise in the North Pacific over a two-week period in Spring 2016 \citep{ribalet2019seaflow}. The data set includes 6,336 scatterplots (referred to as cytograms), each representing the cells measured during a three-minute time interval. We note that clustering is commonly used to distinguish different cell types within a cytogram; however, in this oceanographic setting, the goal is to understand the variability in cytograms across different time points. Thus, we take as input in this example a 6,336-by-6,336 dissimilarity matrix that was computed by \citet{cape2020} based on the earth mover's distance (\citet{rubner2000earth}) between cytograms, using an approach similar to what was proposed in \citet{orlova2016earth}. Earth mover's distance, which is also known as Wasserstein's distance, is a common approach to measuring the distance between two distributions. It imagines these distributions as physical mounds of dirt and measures the distance between these in terms of the minimum amount of dirt-moving needed to transform one into the other. In this example, histogram-approximations of the scatterplots are taken as the distributions on which earth mover's distance is computed.

\begin{figure}[ht!]
\includegraphics[width=4.45in,fbox]{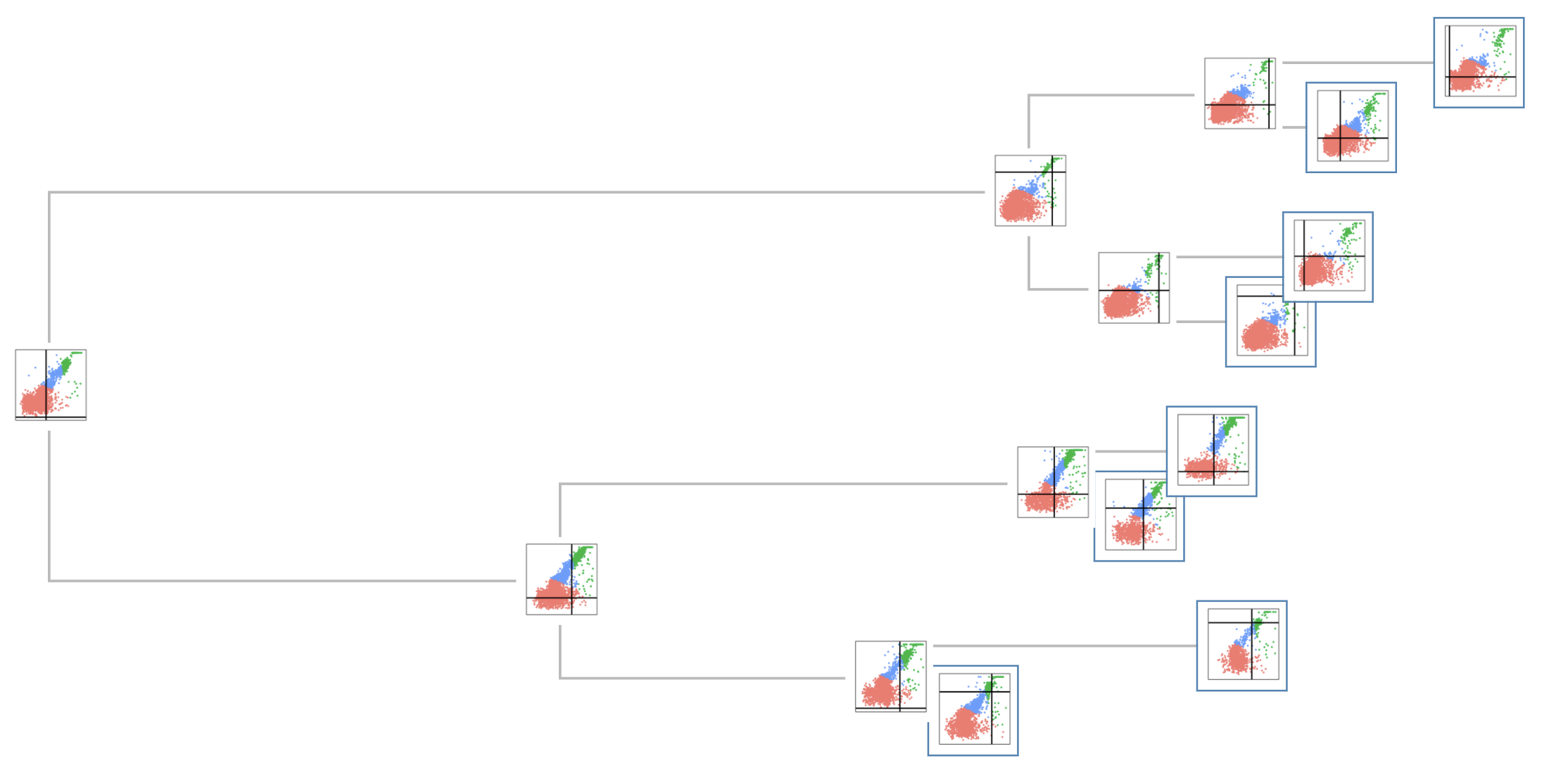} \includegraphics[width=1.55in,fbox]{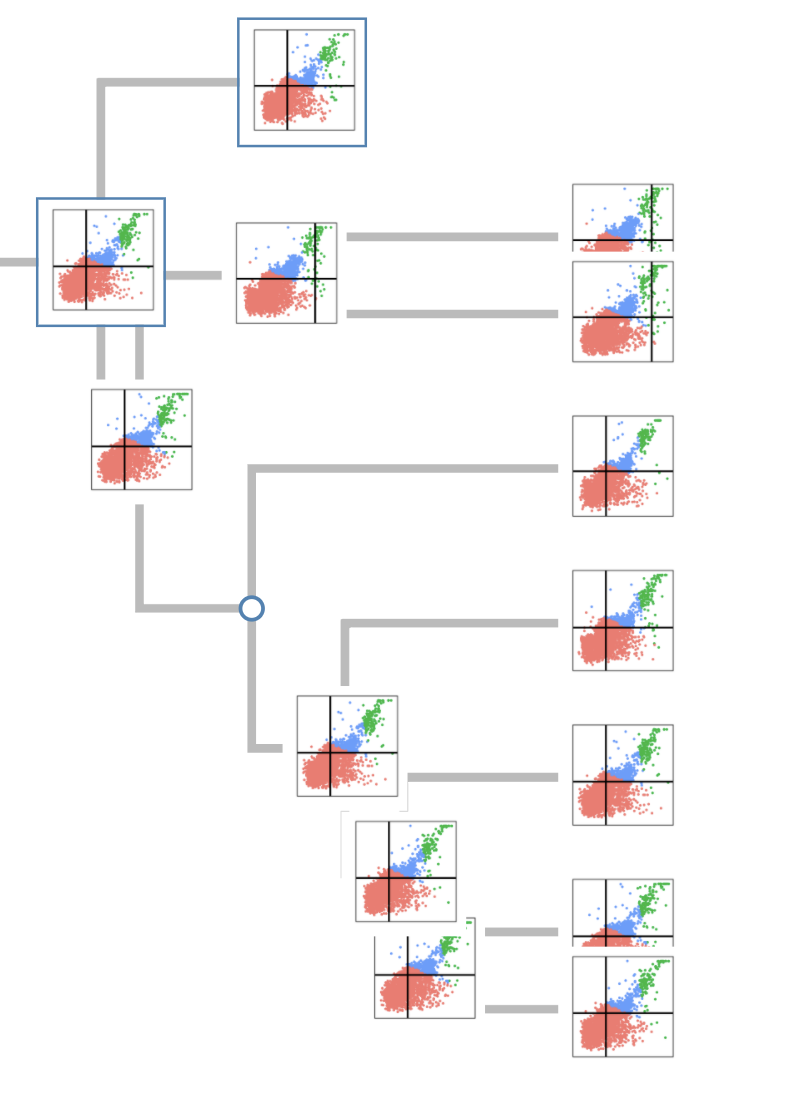} \caption{(Left) Using \texttt{protoshiny} on the SeaFlow data allows one to get a simple high-level view of the general types of structure present in the collection of over 6,000 cytograms before (Right) drilling down into particular branches for more detailed examination.}\label{fig:seaflow}
\end{figure}

Figure \ref{fig:seaflow} shows two screenshots of using \texttt{protoshiny} to explore the dendrogram. In this example, every cytogram is visually labeled by a thumbnail image (showing a two-dimensional projection of the cytogram), and a text label giving the timestamp of the sample shows up on mouseover as a tooltip (not shown in figure). The color of the points corresponds to a crude division (or ``gating'') of the cells into three broad classes of cells. The position of the vertical line represents the date of the measurement (from 2016-04-20 on the far left to 2016-05-04 on the far right), and the position of the horizontal line represents the time of day (from midnight on the bottom to 11:59pm on the top). In the left panel of Figure \ref{fig:seaflow}, one can discern in the three main branches of the dendrogram subtle differences in the point cloud structure, providing a convenient, uncluttered high-level summary of over 6,000 cytograms. Unsurprisingly, exploration of the dendrogram reveals that cytograms whose date-timestamps are close together tend to be clustered together. However, the right panel of Figure \ref{fig:seaflow} shows an exception. From comparing vertical bars in this branch, we observe two cytograms from late in the cruise (2016-05-01) within a branch that otherwise contains cytograms from a single day earlier in the cruise (2016-04-24). Interestingly, the horizontal bar reveals that all cytograms from this branch (across both days) are from the same time of day (10am--11am). Further investigation reveals that the ship was in a very similar latitude at these two dates. The cruise spanned over 16\(^\circ\) of latitude in total while the cytograms in this branch were all within 1.5\(^\circ\) of each other.

The \texttt{protoshiny} package has now been integrated into the SeaFlow data curation pipeline \citep{ribalet2020personal}. In particular, it is used to rapidly check the consistency and correctness of the cell population gating for cytograms by expanding prototype nodes to examine clusters of similar cytograms.

\hypertarget{covid-19}{%
\subsection{COVID-19}\label{covid-19}}

As of April 15, 2022, there have been 503,025,210 confirmed COVID-19 cases and 6,194,288 related deaths recorded worldwide, with 80,576,205 confirmed cases and 988,161 deaths being attributed to the United States \citep{covidcasesjhu}. A positive relationship between human mobility and the spread of the COVID-19 virus has been observed by multiple authors \citep{kraemer2020effect, badr2020association, engle2020staying}. Additionally, there is evidence that COVID-19 has been spread through the air conditioning systems in restaurants \citep{lu2020covid}. We are interested in investigating the mobility patterns in counties in the US, with an additional focus of how this relates to the spread of COVID-19 through the use of hierarchical clustering. We will consider the following three questions:

\begin{enumerate}
\def\labelenumi{\arabic{enumi})}
\tightlist
\item
  Can we cluster US counties by residents' behaviors---i.e., home mobility and restaurant mobility?
\item
  Are there any interesting patterns that emerge from a clustering of US counties by residents' behaviors?
\item
  Do those clustered US counties have similar trajectories of COVID cases?
\end{enumerate}

To address these three questions, we use minimax linkage (as detailed in Section \ref{hierarchical-clustering-with-prototypes}) to obtain a dendrogram with prototypes and then use \texttt{protoshiny} to inspect the clusters for interesting patterns. We use \texttt{protoshiny}'s image labels to address the second question, and the application's ability to quickly change image labels to address the third question within a single session of \texttt{protoshiny}.

\begin{figure}
\includegraphics[width=1\linewidth,fbox]{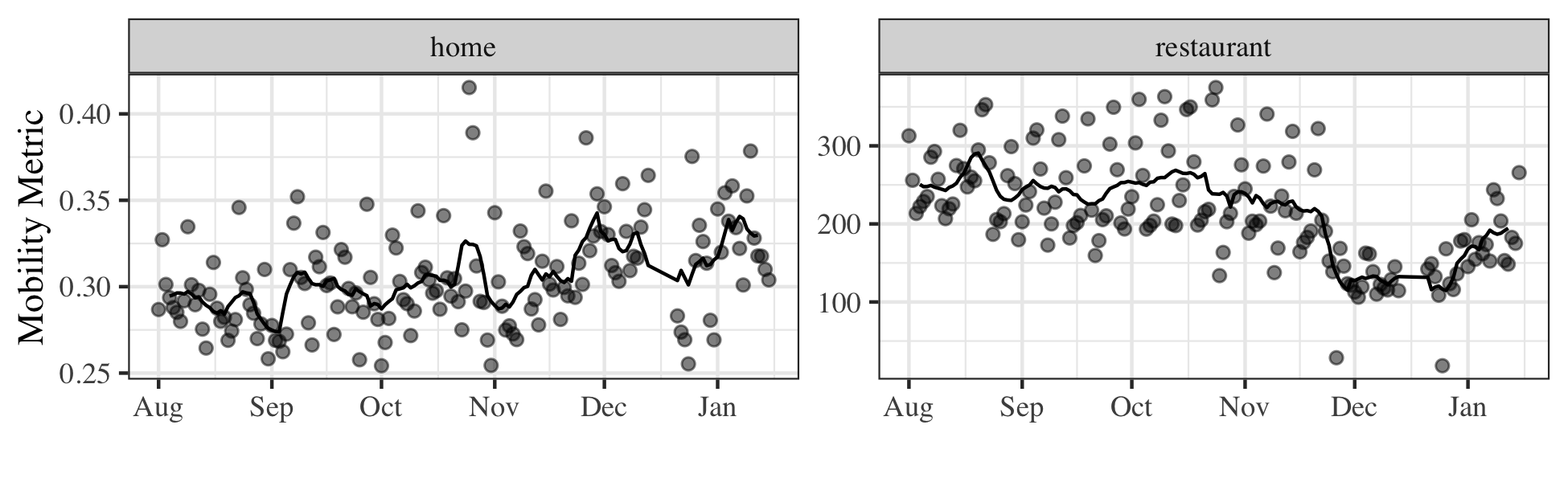} \includegraphics[width=1\linewidth,fbox]{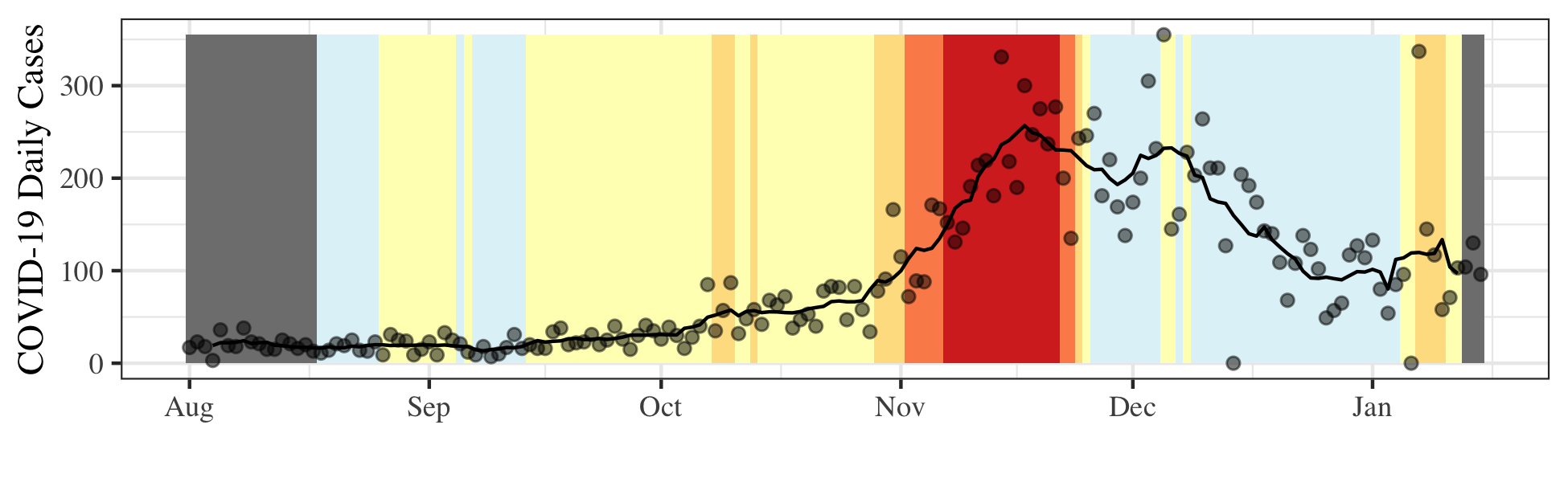} \caption{(Top) Mobility data with a seven-day moving average line for Larimer County, CO from August 1, 2020 to January 15, 2021. While the home mobility measure has an increasing trend over the winter months, there is a clear drop in the restaurant mobility measure over the same time period. (Bottom) COVID-19 case numbers for Larimer County, CO for thee same time period with seven-day moving average on top of the raw data. The background of the plot is colored by trend as compared to two weeks prior. Light blue indicates lower case numbers, yellow indicates unchanged numbers, and orange and red shades indicate increasing levels of case numbers. There is a large spike in cases during the month of November, corresponding to the weeks prior stricter county regulations.}\label{fig:mobility-eda}
\end{figure}

We use two data sources to perform this analysis---mobility data and case numbers. The mobility data come from \citet{safegraph} via the \texttt{covidcast} R package \citep{covidcast}. We pull two mobility metrics from this data source---the fraction of mobile devices that did not leave the immediate area of their home in each day per 100,000 population (home mobility) and the number of daily visits made by those with SafeGraph's apps to restaurants in a certain region per 100,000 population (restaurant mobility). These two mobility metrics are aggregated by county in the US by the CMU Delphi research group as described in their documentation \citep{cmudelphi}, and we have pulled data from August 1, 2020 to Jan 15, 2021. The top of Figure \ref{fig:mobility-eda} shows the mobility data for one such county---Larimer County, CO---for the time period discussed. While the home mobility measure has a slight increase over the winter months, there is a clear drop in the restaurant mobility over the same time period. This can be explained by three potential factors: (1) an increase in restrictions on restaurants limiting indoor dining in the county that began on November 24, 2020 \citep{larimer}, (2) lower temperatures making outdoor dining less pleasant, and (3) the suspension of on-campus learning at Colorado State University on November 30, 2020 \citep{csu-online}. It is of note that indoor dining had an increase in Larimer County in late December, which preceded the loosening of restrictions to allow for indoor dining on January 4, 2021 \citep{larimer-orange} and may correspond to the return of students to the county.

In addition to clustering counties by mobility, we will also look at the COVID-19 case numbers over time in each county. By looking at case numbers clustered by mobility trends, we can hope to gain some insight into the relationship between them and address our third question. The case numbers data is pulled from The New York Times, based on reports from state and local health agencies \citep{covidcases}. The bottom of Figure \ref{fig:mobility-eda} shows the case numbers for Larimer County, CO for the same date range (August 1, 2020 to Jan 15, 2021). There is a large spike in cases during the month of November, corresponding to the weeks prior to the stricter county regulations.

\begin{figure}[t!]
\includegraphics[width=6.25in]{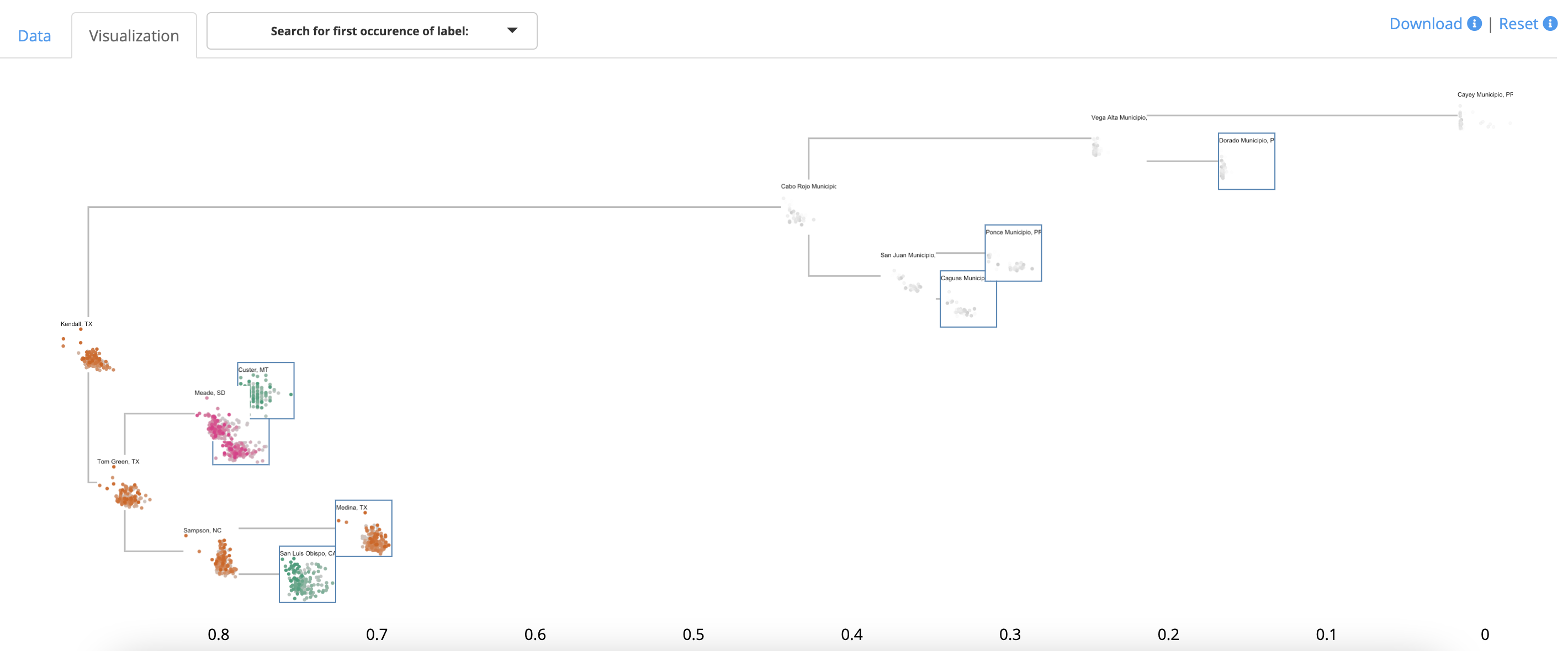} \caption{Using \texttt{protoshiny} on the COVID-19 mobility data before drilling down into particular branches for more detailed examination. The thumbnails are scatterplots of the two mobility metrics colored by region as defined by the 2010 U.S. Census -- gray indicates U.S. territories, pink indicates counties in the Midwest region, orange indicates counties in the South region, purple indicates counties in the Northeast region, and teal indicates counties in the West region -- with the increasing intensity of color to indicate time.}\label{fig:mobility-protoshiny}
\end{figure}

In order to cluster the US counties by mobility, we create a dissimilarity matrix consisting of one minus the correlation of the vectorized mobility data between counties after centering and scaling the individual features, where the ``vectorized'' mobility data refers to stacking the two sets of metrics on top of one another to create a vector of points. To avoid issues with missing data values, we have removed counties that do not have complete mobility data. This results in 2,428 counties to be clustered. Figure \ref{fig:mobility-protoshiny} shows the initial overview of this clustering as seen within \texttt{protoshiny}. The thumbnails are scatterplots of the two mobility metrics colored by region as defined by the 2010 U.S. Census \citep{census2010} with the increasing intensity of color to indicate time. Of the thumbnails displayed in the initial screenshot of Figure \ref{fig:mobility-protoshiny}, gray indicates U.S. territories, pink indicates counties in the Midwest region, orange indicates counties in the South region, purple indicates counties in the Northeast region, and teal indicates counties in the West region.

\begin{figure}[t!]
\includegraphics[width=6.25in,fbox]{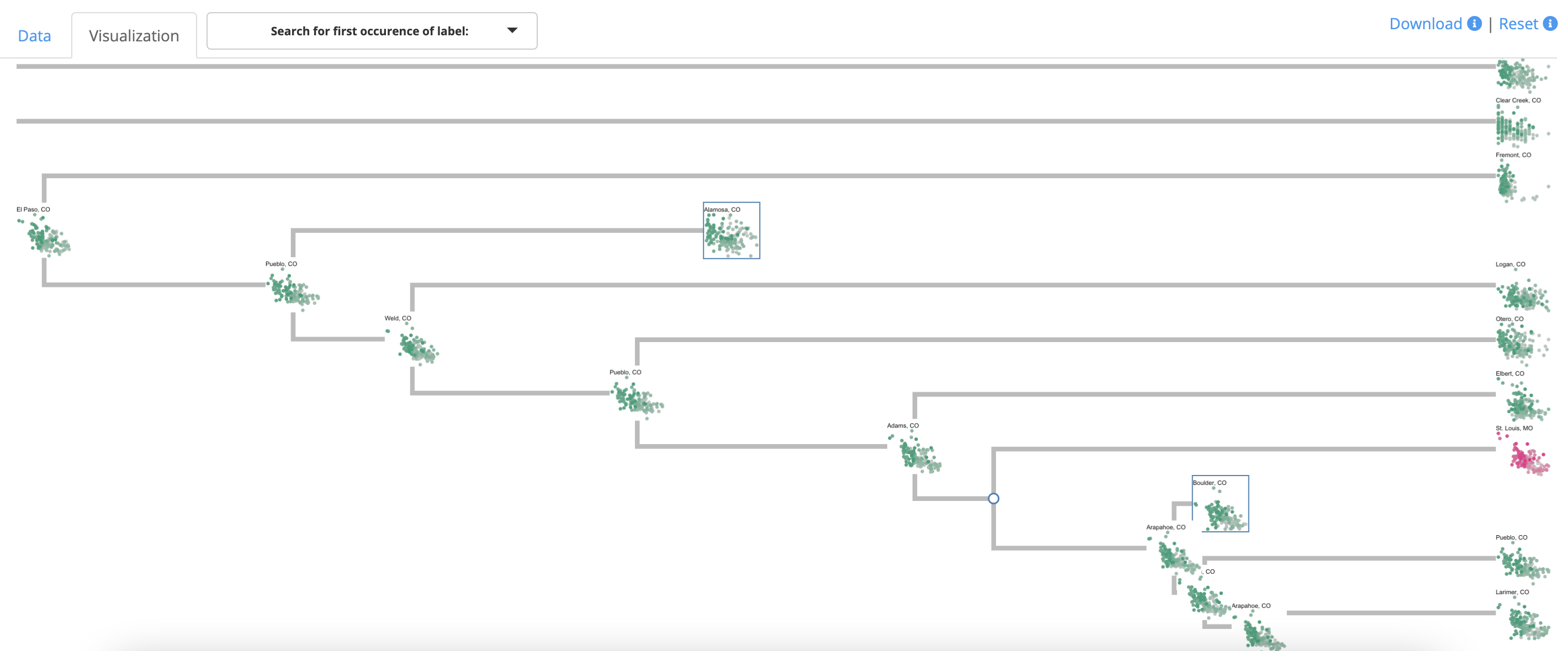} \includegraphics[width=6.25in,fbox]{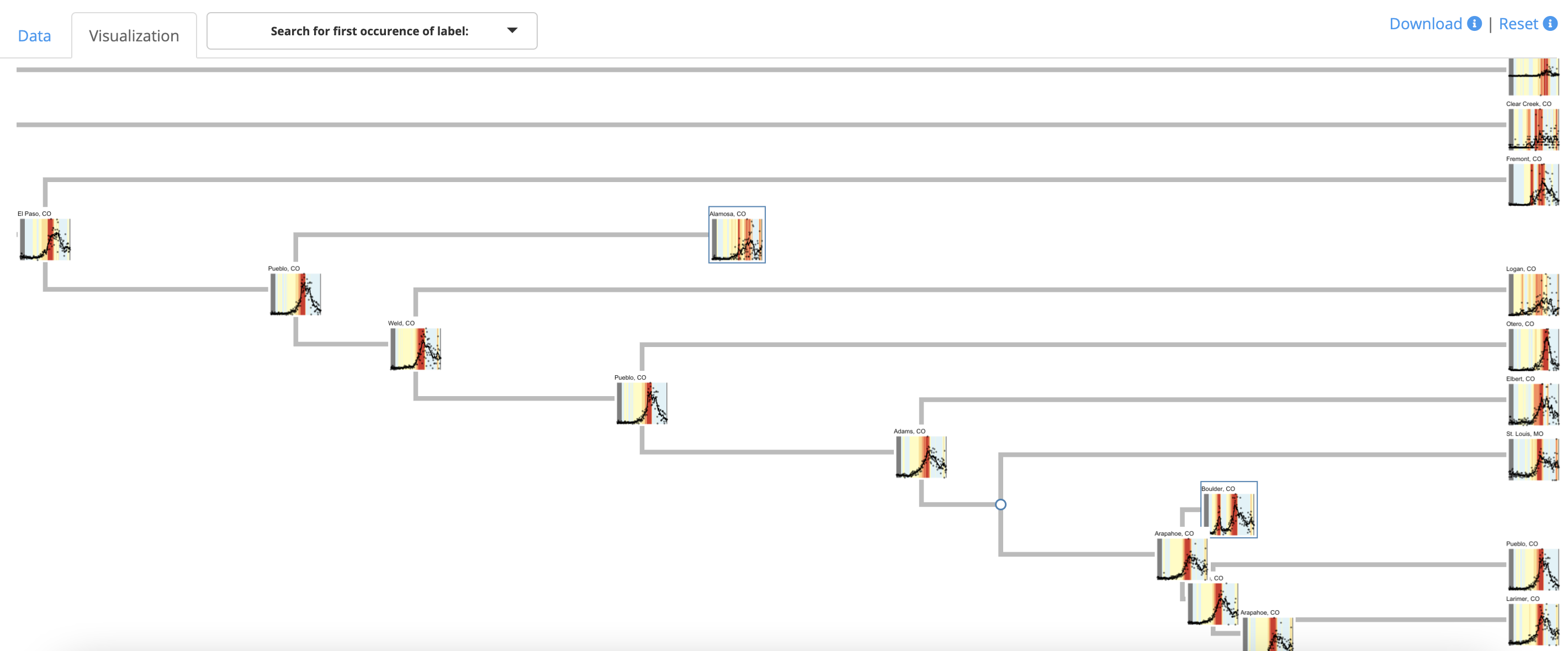} \caption{(Top) Using \texttt{protoshiny} on the COVID-19 mobility data to examine the clusters containing Larimer County, CO. (Bottom) By changing the labels to scatterplots of COVID-19 cases per day, we can explore possible connections between mobility and COVID-19 case numbers in these counties.}\label{fig:mobility-protoshiny-larimer}
\end{figure}

It is straightforward to drill down to the first instance of Larimer County, CO, as seen in the top of Figure \ref{fig:mobility-protoshiny-larimer} by using the search functionality. Interestingly, Larimer County is a prototype for the neighboring Weld County, CO. From this detailed view in \texttt{protoshiny}, we are able to see clear geographic clusters have occurred even though geography was not included in the dissimilarity matrix. Specifically, all of the counties colored in green are in the west region of the US, and, further, they are all Colorado counties. This indicates there are geographic patterns in mobility and distancing behavior, specifically with regards to staying home and going to restaurants. This is not unexpected due to the state-wide policies that have or have not been put in place in each state at different times.

By changing the labels in \texttt{protoshiny} to scatterplots of COVID-19 cases per day without altering the dissimilarity matrix used for clustering (see bottom of Figure \ref{fig:mobility-protoshiny-larimer}), we can explore possible connections between mobility and the COVID-19 case numbers in these counties. In this instance, we have updated the image labels for each prototype without changing the underlying clustering. From this view, the clustered Colorado counties appear to all have a similar pattern, with a large spike in November. The exception to this is Boulder County, which is shown as a prototype for a different cluster than Larimer County containing Boulder and Denver counties. This suggests that the mobility and distancing behavior in Boulder is slightly different than Larimer, which is captured by the clustering based on mobility and distancing, and might explain a portion of the difference in case numbers. This illustrates that a data analyst can change the thumbnails in \texttt{protoshiny} without having to reload the dendrogram, which allows this change to be quickly accomplished for the same clustering object and can lead to further insights into the data.

\hypertarget{discussion}{%
\section{Discussion}\label{discussion}}

In this paper, we have presented a workflow and accompanying tool that renders dendrograms from hierarchical clustering useful for exploring data sets at scales of interest. The approach combines interactivity with the idea of using prototypes to summarize branches of a dendrogram. The result is a novel way to gain insight from hierarchical clustering on more realistically sized data sets. We have also presented three case studies to highlight the multiple strengths of the approach in diverse domains.

In addition to the functionality presented in this paper, \texttt{protoshiny} also features the ability to save and download the resulting clusters that result from a session. It is also possible to load a clustering resulting from a previously saved session in \texttt{protoshiny} back into the tool and display the dendrogram with expanded and contracted branches exactly as before. This is a step towards more reproducible analyses resulting from an interactive online application. The ability to export the current clusters allows for more natural integration of \texttt{protoshiny} into users' current analysis workflows. A related direction for future work will be to add more features to the R API for fast loading of clustering objects and labels into the browser tool.

While \texttt{protoshiny} does expand on the utility of dendrograms for larger data sets, a current limitation of the tool is extreme scalability. One can imagine a massive dendrogram that would not even be loadable into the tool due to the framework in place. Currently, the entire tree is loaded from the server side into the client side of the application at one time and different branches are hidden or shown to the user in their browser via clicks. In the case of a massive tree, it may make sense to only load relevant parts of the tree as a user clicks through and expands branches. While it is entirely possible to shift the framework of \texttt{protoshiny}, a limitation remains in that the clustering must first be calculated on the entire data set. Currently the authors have loaded data sets of size about \(20,000\) observations with no issue. Nonetheless, even in its current form \texttt{protoshiny} provides new capabilities for practitioners to explore their data sets in a way that previously was not possible.

\hypertarget{acknowledgments}{%
\section*{Acknowledgments}\label{acknowledgments}}
\addcontentsline{toc}{section}{Acknowledgments}

This work was supported by grants by the Simons Collaboration on Computational Biogeochemical Modeling of Marine Ecosystems/CBIOMES (Grant ID: 549939 to JB) and NSF CAREER Award DMS-1653017. The authors would like to thank Mattias Rolf Cape and Fran\c{c}ois Ribalet for providing the flow cytometry dissimilarity matrix, as well as Annette Hynes and Fran\c{c}ois Ribalet for information on how \texttt{protoshiny} has been used in the SeaFlow data curation process.

\hypertarget{supplementary-material}{%
\section*{Supplementary Material}\label{supplementary-material}}
\addcontentsline{toc}{section}{Supplementary Material}

The supplementary material contains comparisons to other tools for producing static and interactive dendrograms. Additionally, images used as labels for all examples in the paper and code for producing the clusters are available at \url{https://github.com/andeek/protoshiny-code}. All R data objects used for exploration within \texttt{protoshiny} are available within the package, which is available on CRAN.

\hypertarget{disclosure-statement}{%
\section*{Disclosure Statement}\label{disclosure-statement}}
\addcontentsline{toc}{section}{Disclosure Statement}

The authors report there are no competing interests to declare.

\bibliographystyle{agsm}
\bibliography{tas_files/refs.bib}

@article{rubner2000earth,
  title={The earth mover's distance as a metric for image retrieval},
  author={Rubner, Yossi and Tomasi, Carlo and Guibas, Leonidas J},
  journal={International journal of computer vision},
  volume={40},
  number={2},
  pages={99--121},
  year={2000},
  publisher={Springer}
}

@book{hastie2009elements,
  title={The elements of statistical learning: data mining, inference, and prediction},
  author={Hastie, Trevor and Tibshirani, Robert and Friedman, Jerome H and Friedman, Jerome H},
  volume={2},
  year={2009},
  publisher={Springer}
}

@article{swalwell2011seaflow,
  title={SeaFlow: A novel underway flow-cytometer for continuous observations of phytoplankton in the ocean},
  author={Swalwell, Jarred E and Ribalet, Francois and Armbrust, E Virginia},
  journal={Limnology and Oceanography: Methods},
  volume={9},
  number={10},
  pages={466--477},
  year={2011},
  publisher={Wiley Online Library}
}

@article{ribalet2019seaflow,
  title={SeaFlow data v1, high-resolution abundance, size and biomass of small phytoplankton in the North Pacific},
  author={Ribalet, Fran{\c{c}}ois and Berthiaume, Chris and Hynes, Annette and Swalwell, Jarred and Carlson, Michael and Clayton, Sophie and Hennon, Gwenn and Poirier, Camille and Shimabukuro, Eric and White, Angelicque and Armbrust, E. Virginia},
  journal={Scientific data},
  volume={6},
  number={1},
  pages={1--8},
  year={2019},
  publisher={Nature Publishing Group}
}

@conference{cape2020, 
  author = {Cape, Mattias Rolf and Ribalet, Fran{\c{c}}ois and Bien, Jacob and Hyun, Sangwon and Armbrust, E. Virginia
}, 
  title = {OB14F-0437 - Determining ecological provinces from optical cytometric data in the North Pacific Ocean}, 
  booktitle = {Ocean Sciences Meeting}, 
  year = {2020}, 
  address = {San Diego, CA}, 
  month = {Feb. 17}, 
  url = {https://agu.confex.com/agu/osm20/meetingapp.cgi/Paper/657891} 
}

@article{orlova2016earth,
  title={Earth mover’s distance (EMD): a true metric for comparing biomarker expression levels in cell populations},
  author={Orlova, Darya Y and Zimmerman, Noah and Meehan, Stephen and Meehan, Connor and Waters, Jeffrey and Ghosn, Eliver EB and Filatenkov, Alexander and Kolyagin, Gleb A and Gernez, Yael and Tsuda, Shanel and Moore, Wayne and Moss, Richard B and Herzenberg, Leonore A and Walther, G},
  journal={PloS one},
  volume={11},
  number={3},
  pages={e0151859},
  year={2016},
  publisher={Public Library of Science San Francisco, CA USA}
}

@misc{ribalet2020personal,
  author = "Ribalet, Fran{\c{c}}ois and Hynes, Annette",
  note = "Received: August 12, 2020",
  year = "2020",
  howpublished = "Personal Communication"
}

@article{vig2012tag,
  title={The tag genome: Encoding community knowledge to support novel interaction},
  author={Vig, Jesse and Sen, Shilad and Riedl, John},
  journal={ACM Transactions on Interactive Intelligent Systems (TiiS)},
  volume={2},
  number={3},
  pages={1--44},
  year={2012},
  publisher={ACM New York, NY, USA}
}

@article{harper2015movielens,
  title={The movielens datasets: History and context},
  author={Harper, F Maxwell and Konstan, Joseph A},
  journal={Acm transactions on interactive intelligent systems (tiis)},
  volume={5},
  number={4},
  pages={1--19},
  year={2015},
  publisher={ACM New York, NY, USA}
}

@article{bien2011hierarchical,
  title={Hierarchical clustering with prototypes via minimax linkage},
  author={Bien, Jacob and Tibshirani, Robert},
  journal={Journal of the American Statistical Association},
  volume={106},
  number={495},
  pages={1075--1084},
  year={2011},
  publisher={Taylor \& Francis}
}

@Manual{shiny,
  title = {shiny: Web Application Framework for R},
  author = {Winston Chang and Joe Cheng and JJ Allaire and Yihui Xie and Jonathan McPherson},
  year = {2017},
  note = {R package version 1.0.5},
  url = {https://CRAN.R-project.org/package=shiny},
}

@article{bostock2011d3,
 author = {Bostock, Michael and Ogievetsky, Vadim and Heer, Jeffrey},
 title = {D3 Data-Driven Documents},
 journal = {IEEE Transactions on Visualization and Computer Graphics},
 issue_date = {December 2011},
 volume = {17},
 number = {12},
 month = dec,
 year = {2011},
 issn = {1077-2626},
 pages = {2301--2309},
 numpages = {9},
 url = {http://dx.doi.org/10.1109/TVCG.2011.185},
 doi = {10.1109/TVCG.2011.185},
 acmid = {2068631},
 publisher = {IEEE Educational Activities Department},
 address = {Piscataway, NJ, USA},
}

@article{ao2005clustag,
author = {Ao, S. I. and Yip, Kevin and Ng, Michael and Cheung, David and Fong, Pui-Yee and Melhado, Ian and Sham, Pak C.},
title = {CLUSTAG: hierarchical clustering and graph methods for selecting tag SNPs},
journal = {Bioinformatics},
volume = {21},
number = {8},
pages = {1735-1736},
year = {2005},
}

@misc{asavideolibrary,
  title = {{ASA Sections on: Statistical Computing Statistical Graphics Video Library}},
  author = {{ASA Sections on: Statistical Computing Statistical Graphics}},
  howpublished = {\url{http://stat-graphics.org/movies/}},
  note = {Accessed: 2018-09-26},
  year = {2018}
}

@Manual{plotly,
  title = {plotly for R},
  author = {Carson Sievert},
  year = {2018},
  url = {https://plotly-book.cpsievert.me},
}

@Manual{ggvis,
  title = {{ggvis: Interactive Grammar of Graphics}},
  author = {Winston Chang and Hadley Wickham},
  year = {2016},
  note = {R package version 0.4.3},
  url = {https://CRAN.R-project.org/package=ggvis},
  }

@Manual{animint,
    title = {animint: Interactive animations},
    author = {Toby Dylan Hocking and Susan VanderPlas and Carson Sievert and Kevin Ferris and Tony Tsai and Faizan Khan},
    note = {R package version 2017.01.04},
    url = {https://github.com/tdhock/animint},
    year = {2017}
  }

@Article{Swayne1999,
author="Swayne, Deborah F. and Klinke, Sigbert",
title="Introduction to the special issue on interactive graphical data analysis: What is interaction?",
journal="Computational Statistics",
year="1999",
volume="14",
number="1",
pages="1--6",
issn="0943-4062",
}

@article{kaplan2017interactive,
  title={An interactive graphical method for community detection in network data},
  author={Kaplan, Andee and Hofmann, Heike and Nordman, Daniel},
  journal={Computational Statistics},
  volume={32},
  number={2},
  pages={535--557},
  year={2017},
  publisher={Springer}
}

@article{kaplan2016putting,
  title={Putting down roots: a graphical exploration of community attachment},
  author={Kaplan, Andee J and Hare, Eric R},
  journal={Computational Statistics},
  volume={34},
  number={4},
  pages={1449--1464},
  year={2019},
  publisher={Springer}
}

@inproceedings{sievert2014ldavis,
  title={LDAvis: A method for visualizing and interpreting topics},
  author={Sievert, Carson and Shirley, Kenneth},
  booktitle={Proceedings of the workshop on interactive language learning, visualization, and interfaces},
  pages={63--70},
  year={2014}
}

@Manual{protoclust,
    title = {protoclust: Hierarchical Clustering with Prototypes},
    author = {Jacob Bien and Rob Tibshirani},
    year = {2017},
    note = {R package version 1.6.1},
  }

@article{langfelder2007defining,
    author = {Langfelder, Peter and Zhang, Bin and Horvath, Steve},
    title = "{Defining clusters from a hierarchical cluster tree: the Dynamic Tree Cut package for R}",
    journal = {Bioinformatics},
    volume = {24},
    number = {5},
    pages = {719-720},
    year = {2007},
    month = {11},
    issn = {1367-4803},
}

@Manual{dynamicTreeCut,
    title = {dynamicTreeCut: Methods for Detection of Clusters in Hierarchical Clustering Dendrograms},
    author = {Peter Langfelder and Bin Zhang and Steve Horvath},
    year = {2016},
    note = {R package version 1.63-1},
    url = {https://CRAN.R-project.org/package=dynamicTreeCut},
  }

@article{swayne1998xgobi,
  title={XGobi: Interactive dynamic data visualization in the X Window System},
  author={Swayne, Deborah F and Cook, Dianne and Buja, Andreas},
  journal={Journal of computational and Graphical Statistics},
  volume={7},
  number={1},
  pages={113--130},
  year={1998},
  publisher={Taylor \& Francis Group}
}

@article{dang2010stacking,
  title={Stacking graphic elements to avoid over-plotting},
  author={Dang, Tuan Nhon and Wilkinson, Leland and Anand, Anushka},
  journal={IEEE Transactions on Visualization and Computer Graphics},
  volume={16},
  number={6},
  pages={1044--1052},
  year={2010},
  publisher={IEEE}
}

@inproceedings{cottam2013overplotting,
  title={Overplotting: Unified solutions under abstract rendering},
  author={Cottam, Joseph and Lumsdaine, Andrew and Wang, Peter},
  booktitle={2013 IEEE International Conference on Big Data},
  pages={9--16},
  year={2013},
  organization={IEEE}
}

@Manual{covidcast,
    title = {covidcast: Client for Delphi's COVIDcast API},
    author = {Jacob Bien and Logan Brooks and David Farrow and Alex Reinhart and Ryan Tibshirani},
    year = {2021},
    note = {https://cmu-delphi.github.io/covidcast/covidcastR/,
https://github.com/cmu-delphi/covidcast},
  }

@misc{safegraph,
  title = {SafeGraph},
  howpublished = {\url{https://www.safegraph.com}},
  note = {Accessed: 2021-01-25},
  year = {2021}
}

@misc{larimer,
  title = {{Larimer County to Move to Safer at Home Level Red of Colorado's Dial}},
  author = {Katie O'Donnell},
  year = {2020},
  howpublished = {\url{https://www.larimer.org/spotlights/2020/11/20/larimer-county-move-safer-home-level-red-colorado-s-dial}},
  note = {Accessed: 2021-01-25}
}

@misc{covidcases,
  title = {{Coronavirus (Covid-19) Data in the United States}},
  author = {{The New York Times}},
  year = {2021},
  howpublished = {\url{https://github.com/nytimes/covid-19-data}},
  note = {Accessed: 2021-01-15}
}

@article{engle2020staying,
  title={Staying at home: mobility effects of covid-19},
  author={Engle, Samuel and Stromme, John and Zhou, Anson},
  journal={Available at SSRN},
  year={2020}
}

@article{badr2020association,
  title={Association between mobility patterns and COVID-19 transmission in the USA: a mathematical modelling study},
  author={Badr, Hamada S and Du, Hongru and Marshall, Maximilian and Dong, Ensheng and Squire, Marietta M and Gardner, Lauren M},
  journal={The Lancet Infectious Diseases},
  volume={20},
  number={11},
  pages={1247--1254},
  year={2020},
  publisher={Elsevier}
}

@article{kraemer2020effect,
  title={The effect of human mobility and control measures on the COVID-19 epidemic in China},
  author={Kraemer, Moritz UG and Yang, Chia-Hung and Gutierrez, Bernardo and Wu, Chieh-Hsi and Klein, Brennan and Pigott, David M and Du Plessis, Louis and Faria, Nuno R and Li, Ruoran and Hanage, William P and others},
  journal={Science},
  volume={368},
  number={6490},
  pages={493--497},
  year={2020},
  publisher={American Association for the Advancement of Science}
}

@article{lu2020covid,
  title={COVID-19 outbreak associated with air conditioning in restaurant, Guangzhou, China, 2020},
  author={Lu, Jianyun and Gu, Jieni and Li, Kuibiao and Xu, Conghui and Su, Wenzhe and Lai, Zhisheng and Zhou, Deqian and Yu, Chao and Xu, Bin and Yang, Zhicong},
  journal={Emerging infectious diseases},
  volume={26},
  number={7},
  pages={1628},
  year={2020},
  publisher={Centers for Disease Control and Prevention}
}

@misc{covidcasesjhu,
  title = {{Coronavirus Resource Center}},
  author = {{Johns Hopkins University \& Medicine}},
  year = {2022},
  howpublished = {\url{https://coronavirus.jhu.edu}},
  note = {Accessed: 2022-04-15}
}

@misc{larimer-orange,
  title = {{Safer at Home Level Orange}},
  author = {{Larimer County}},
  year = {2021},
  howpublished = {\url{https://www.larimer.org/health/communicable-disease/coronavirus-covid-19/safer-at-home}},
  note = {Accessed: 2021-02-08}
}

@misc{groliers,
  title = {{Data for MATLAB Hackers}},
  author = {Sam Roweis},
  howpublished = {\url{https://cs.nyu.edu/~roweis/data.html}},
  note = {Accessed: 2021-02-16},
  year = {2008}
}

@Manual{rstats,
    title = {R: A Language and Environment for Statistical Computing},
    author = {{R Core Team}},
    organization = {R Foundation for Statistical Computing},
    address = {Vienna, Austria},
    year = {2021},
    url = {https://www.R-project.org/},
  }

@misc{csu-online,
  title = {{Fall 2020 Framework}},
  author = {{Colorado State University}},
  year = {2020},
  howpublished = {\url{https://covid.colostate.edu/kb/fall-2020-framework/}},
  note = {Accessed: 2021-03-02}
}

@Manual{devtools,
    title = {devtools: Tools to Make Developing R Packages Easier},
    author = {Hadley Wickham and Jim Hester and Winston Chang},
    year = {2020},
    note = {R package version 2.3.2},
    url = {https://CRAN.R-project.org/package=devtools},
  }

@misc{census2010,
  author = {{U.S. Census Bureau}},
  title = {{2010 Census Regions and Divisions of the United States}},
  year = {2010},
  howpublished = {\url{https://www2.census.gov/geo/pdfs/maps-data/maps/reference/us_regdiv.pdf}},
  note = {Accessed: 2021-04-13}
}

@article{sieger2017interactive,
   author = {Tomáš Sieger and Catherine B. Hurley and Karel Fišer and Claudia Beleites},
   title = {Interactive Dendrograms: The R Packages idendro and idendr0},
   journal = {Journal of Statistical Software, Articles},
   volume = {76},
   number = {10},
   year = {2017},
   issn = {1548-7660},
   pages = {1--22},
}

@Manual{collapsibleTree,
    title = {collapsibleTree: Interactive Collapsible Tree Diagrams using `D3.js'},
    author = {Adeel Khan},
    year = {2018},
    note = {R package version 0.1.7},
    url = {https://CRAN.R-project.org/package=collapsibleTree},
  }

@Manual{palmerpenguins,
    title = {palmerpenguins: Palmer Archipelago (Antarctica) penguin data},
    author = {Allison Marie Horst and Alison Presmanes Hill and Kristen B Gorman},
    year = {2020},
    note = {R package version 0.1.0},
    url = {https://allisonhorst.github.io/palmerpenguins/},
  }

@book{tukey1977exploratory,
  title={Exploratory data analysis},
  author={Tukey, John W and others},
  volume={2},
  year={1977},
  publisher={Reading, MA}
}

@book{theus2008interactive,
  title={Interactive graphics for data analysis: principles and examples},
  author={Theus, Martin and Urbanek, Simon},
  year={2008},
  publisher={CRC Press}
}

@article{satyanarayan2016vega,
  title={Vega-lite: A grammar of interactive graphics},
  author={Satyanarayan, Arvind and Moritz, Dominik and Wongsuphasawat, Kanit and Heer, Jeffrey},
  journal={IEEE transactions on visualization and computer graphics},
  volume={23},
  number={1},
  pages={341--350},
  year={2016},
  publisher={IEEE}
}

@article{su2017glimma,
  title={Glimma: interactive graphics for gene expression analysis},
  author={Su, Shian and Law, Charity W and Ah-Cann, Casey and Asselin-Labat, Marie-Liesse and Blewitt, Marnie E and Ritchie, Matthew E},
  journal={Bioinformatics},
  volume={33},
  number={13},
  pages={2050--2052},
  year={2017},
  publisher={Oxford University Press}
}

@book{young2011visual,
  title={Visual statistics: seeing data with dynamic interactive graphics},
  author={Young, Forrest W and Valero-Mora, Pedro M and Friendly, Michael},
  year={2011},
  publisher={John Wiley \& Sons}
}

@article{friedman2002john,
  title={John W. Tukey's work on interactive graphics},
  author={Friedman, Jerome H and Stuetzle, Werner},
  journal={The Annals of Statistics},
  volume={30},
  number={6},
  pages={1629--1639},
  year={2002},
  publisher={Institute of Mathematical Statistics}
}

@article{unwin1996interactive,
  title={Interactive graphics for data sets with missing values—MANET},
  author={Unwin, Antony and Hawkins, George and Hofmann, Heike and Siegl, Bernd},
  journal={Journal of Computational and Graphical Statistics},
  volume={5},
  number={2},
  pages={113--122},
  year={1996},
  publisher={Taylor \& Francis}
}

@article{zhao2005hierarchical,
  title={Hierarchical clustering algorithms for document datasets},
  author={Zhao, Ying and Karypis, George and Fayyad, Usama},
  journal={Data mining and knowledge discovery},
  volume={10},
  number={2},
  pages={141--168},
  year={2005},
  publisher={Springer}
}

@inproceedings{cutting2017scatter,
  title={Scatter/gather: A cluster-based approach to browsing large document collections},
  author={Cutting, Douglass R and Karger, David R and Pedersen, Jan O and Tukey, John W},
  booktitle={ACM SIGIR Forum},
  volume={51},
  number={2},
  pages={148--159},
  year={2017},
  organization={ACM New York, NY, USA}
}

@article{sorlie2003repeated,
  title={Repeated observation of breast tumor subtypes in independent gene expression data sets},
  author={S{\o}rlie, Therese and Tibshirani, Robert and Parker, Joel and Hastie, Trevor and Marron, James Stephen and Nobel, Andrew and Deng, Shibing and Johnsen, Hilde and Pesich, Robert and Geisler, Stephanie and others},
  journal={Proceedings of the national academy of sciences},
  volume={100},
  number={14},
  pages={8418--8423},
  year={2003},
  publisher={National Acad Sciences}
}

@article{studdert1974balance,
  title={The balance of Roger de Piles: a statistical analysis},
  author={Studdert-Kennedy, W Gerald and Davenport, Michael},
  journal={The Journal of Aesthetics and Art Criticism},
  volume={32},
  number={4},
  pages={493--502},
  year={1974},
  publisher={JSTOR}
}

@article{kigerl2020behind,
  title={Behind the scenes of the underworld: hierarchical clustering of two leaked carding forum databases},
  author={Kigerl, Alex},
  journal={Social Science Computer Review},
  pages={0894439320924735},
  year={2020},
  publisher={SAGE Publications Sage CA: Los Angeles, CA}
}

@article{saint2003convergence,
  title={Convergence or resilience? A hierarchical cluster analysis of the welfare regimes in advanced countries},
  author={Saint-Arnaud, S{\'e}bastien and Bernard, Paul},
  journal={Current sociology},
  volume={51},
  number={5},
  pages={499--527},
  year={2003},
  publisher={Sage Publications}
}

@article{thrun2021distance,
  title={Distance-based clustering challenges for unbiased benchmarking studies},
  author={Thrun, Michael C},
  journal={Scientific reports},
  volume={11},
  number={1},
  pages={1--12},
  year={2021},
  publisher={Nature Publishing Group}
}

@ARTICLE{1016905,
  author={Jinwook Seo and Shneiderman, B.},
  journal={Computer}, 
  title={Interactively exploring hierarchical clustering results [gene identification]}, 
  year={2002},
  volume={35},
  number={7},
  pages={80-86},
  doi={10.1109/MC.2002.1016905}}

@misc{cmudelphi,
  author = {{CMU Delphi Research Group}},
  title = {{Delphi Epidata API SafeGraph}},
  year = {2020},
  howpublished = {\url{https://cmu-delphi.github.io/delphi-epidata/api/covidcast-signals/safegraph.html}},
  note = {Accessed: 2021-04-13}
}

\end{document}